\documentclass[11pt]{article}
\usepackage{graphicx}
\usepackage{latexsym}
\usepackage{amsfonts}
\usepackage{graphicx}
\usepackage{epic}

\def\be{\begin{equation}}
\def\ee{\end{equation}}
\def\beq{\begin{eqnarray}}
\def\eeq{\end{eqnarray}}

\newsavebox{\uuunit}
\sbox{\uuunit}
    {\setlength{\unitlength}{0.825em}
     \begin{picture}(0.6,0.7)
        \thinlines
        \put(0,0){\line(1,0){0.5}}
        \put(0.15,0){\line(0,1){0.7}}
        \put(0.35,0){\line(0,1){0.8}}
       \multiput(0.3,0.8)(-0.04,-0.02){12}{\rule{0.5pt}{0.5pt}}
     \end {picture}}
\newcommand {\unity}{\mathord{\!\usebox{\uuunit}}}

\begin{document}

\begin{titlepage}

\begin{flushright}
ULB-TH/04-05\\

\end{flushright}
\vskip 1.0cm

\begin{centering}

{\large {\bf Hyperbolic Kac Moody
Algebras and Einstein Billiards}}

\vspace{.5cm}

Sophie de Buyl\footnote{Aspirant du Fonds National de la Recherche
Scientifique, Belgique}, and Christiane Schomblond
\\
\vspace{.7cm} {\small Physique Th\'eorique et Math\'ematique and\\
International Solvay Institutes,\\
Universit\'e Libre
de Bruxelles, \\ C.P. 231, B-1050, Bruxelles, Belgium.\\sdebuyl@ulb.ac.be,
 cschomb@ulb.ac.be}

\vspace{.5cm}

\end{centering}

\begin{abstract}
We identify the hyperbolic Kac Moody algebras for which there
exists a Lagrangian of gravity, dilatons and $p$-forms which
produces a billiard that can be identified with their fundamental
Weyl chamber. Because of the invariance of the billiard upon
toroidal dimensional reduction, the list of admissible algebras is
determined by the existence of a Lagrangian in three space-time
dimensions, where a systematic analysis can be carried out since
only zero-forms are involved. We provide all highest
dimensional parent Lagrangians with their full spectrum of
$p$-forms and dilaton couplings. We confirm, in particular, that
for the rank 10 hyperbolic algebra, $CE_{10} =
A_{15}^{(2)\wedge}$, also known as the dual of
$B_8^{\wedge\wedge}$, the maximally oxidized Lagrangian is $9$
dimensional and involves besides gravity, 2 dilatons, a $2$-form,
a $1$-form and a $0$-form.
\end{abstract}

\vfill
\end{titlepage}

\section{Introduction}

It has been shown recently that the dynamics of the gravitational
scale factors becomes equivalent, in the vicinity of a spacelike
singularity, to that of a relativistic particle moving freely on
an hyperbolic billiard and bouncing on its walls
\cite{BKL}-\cite{DHN}. A criterion for the gravitational dynamics
to be chaotic is that the billiard has a finite volume. This in
turn stems from the remarkable property that the billiard can be
identified with the fundamental Weyl chamber of an hyperbolic Kac Moody
algebra. Some of these algebras are well known: in particular, the
famous hyperbolic algebras\footnote{$E_{10} = E_8^{\wedge
\wedge}, BE_{10}= B_8^{\wedge\wedge}, DE_{10} = D_8^{\wedge\wedge}, AE_n =
A_{n-2}^{\wedge\wedge}$; more generally, the names here given to
the algebras are taken from \cite{DdBHS} and \cite{HJ}; in the
table of the Dynkin diagrams given in \cite{HJ}, the name
$D_{r+1}^{(2)}$ should be replaced by $D_{r+1}^{(2)\wedge}$.}
$E_{10}, BE_{10}, DE_{10}$ \cite{DH2, Fre, BGH} are
related to strings, supergravities and M-theory; the
$AE_{n}, n<10$
\cite{DHJN} emerge from pure gravity in various dimensions and
more generally, the algebras that are overextensions of finite
dimensional simple Lie algebras \cite{CJLP, DdBHS} - also twisted
overextensions \cite{HJ} - are associated to gravitational models
that reduce to $G/H$ coset models upon toroidal dimensional
reduction to $D=3$. Several other hyperbolic algebras also appear
in the billiard analysis of $D=4$ and $D=5$ spatially homogeneous
cosmological models \cite{dBPS}. This kind of analysis has
attracted a lot of interest recently in connection with
$U$-dualities \cite{BP} and hidden symmetries of $M$-theory
\cite{BJ1,PW,BJall1,DHN2,BJall2,En}.
\newline

The purpose of this paper is twofold: first we select all
hyperbolic Kac Moody algebras for which a billiard description
exists and then we explicitly construct all Lagrangians describing
gravity coupled to dilatons and $p$-forms producing these
billiards.
\newline

We are able to give exhaustive results because (i) the hyperbolic
algebras are all known and classified\footnote{Note however six
missing cases in \cite{S}, two with rank 3, two with rank 4 and
two with rank 5; their Dynkin diagrams are displayed at the end of
the paper.} \cite{S}, and (ii) only the finite number of algebras
with rank $r$ between $3$ and $10$ are relevant in this context.
Note that there are infinitely many hyperbolic algebras of rank
two and that there exists no hyperbolic algebra of rank $r>10$.
The analysis is considerably simplified because of the invariance
of the billiard under toroidal dimensional reduction to dimensions $D\geq 3$.
Indeed, as explained in \cite{DdBHS}, the billiard region stays
the same, but a symmetry wall in $D$ dimensions may become an
electric or magnetic $p$-form wall in a lower dimension. The
invariance under dimensional reduction implies in particular that
the selection of algebras with a billiard description can be
performed by analyzing Lagrangians in $D=3$ dimensions.
\newline

Simplifications
in $D=3$ occur because only $0$-forms are present: indeed, via
appropriate dualizations, all
$p$-forms can be reduced to $0$-forms. To be concrete, for the
hyperbolic algebras of real rank $r$ between
$3$ and $6$, we first try to reproduce their Dynkin diagram with a
set of $r$ dominant walls comprising one symmetry wall
($\beta^2-\beta^1)$ and
$(r-1)$ scalar walls. If this can be
done, we still have to check that the remaining walls are
subdominant i.e., that they can be written as linear combinations of
the dominant ones with positive coefficients. In particular, this
analysis requires that any dominant set necessarily involves one
magnetic wall and
$(r-2)$ electric walls. Note that our search for gravitational
Lagrangians in $D=3$ is systematic although no symmetry is required.
To deal with the hyperbolic algebras of ranks
$7$ to $10$, it is actually not necessary to first reduce to 3
dimensions: the overextensions of finite simple Lie
algebras have already been associated to billiards of some
Lagrangians and for the remaining four algebras, the rules we have
found in the previous cases allow to straightforwardly construct the
Lagrangian in the maximal oxidation dimension.
\newline

We then analyze which three-dimensional system admits parents in
higher dimensions and construct the Lagrangian in the maximal
oxidation dimension. In order to do so, we take an algebra in the
previous list and we determine successively the maximal spacetime
dimension, the dilaton number, the $p$-form content and the
dilaton couplings:

\begin{enumerate}
\item One considers the Dynkin diagram of the selected algebra and
looks at the length of its "A-chain"\footnote{An "A-chain" of
length $k$ is a chain of $k$ vertices with norm squared equal to
$2$ and simply laced.}, starting with the symmetry root. Our
analysis produces the following oxidation rule: if the A-chain has
length $k$, the theory can be oxidized up to

\begin{enumerate}
\item{$D_{max}= k+2$}, if the next connected root has a norm
squared smaller than
$2$,
\item{$D_{max}=k+1$}, if the next connected root has a norm squared greater
than
$2$.
\end{enumerate}

This generalizes the oxidation rule by \cite{dBHJP} and
\cite{Keur1,Keur2,Keur3}, obtained by group theoretical arguments
applied to coset models.
\newline

\item For given space dimension $d=D-1$ and rank $r$ of the
algebra, the number of dilatons is given by $N=r-d$ because the
dominant walls are required to be $r$ independent linear forms in
the $d$ scale factors $\{\beta^1,...,\beta^d\}$ and the $N$
dilatons.
\newline

\item Because it is known how the $p$-form walls connect
to the A-chain \cite{DdBHS}, one can read on the Dynkin diagram
which $p$-forms\footnote{or their dual $(d-p-1)$-forms} appear in the
maximal oxidation dimension.
\newline

\item The dilaton couplings of the $p$-forms are computed
from the norms and scalar products of the walls which have to
generate the Cartan matrix of the hyperbolic algebra. This means
in particular that, even if the nature of the walls changes during
the oxidation procedure, their norms and scalar products remain
unchanged. Note also that in all dimensions $D>3$ the subdominant
conditions are always satisfied.

\end{enumerate}

As a byproduct of our analysis, we note that, for each billiard
identifiable as the fundamental Weyl chamber of an hyperbolic
algebra, the positive linear combinations of the dominant walls
representing the subdominant ones only contain integer
coefficients. Hence, the dominant walls of the Lagrangian correspond to
the simple roots of the hyperbolic algebra, while the subdominant
ones correspond to non simple positive roots. The gravitational theory does
not give all the positive roots; even the 3-dimensional scalar Lagrangians do
not describe coset spaces in general. Nevertheless, the reflections
relative to the simple roots generate the Weyl group of the
hyperbolic algebra; this group in turn gives an access to other
positive roots and suggests that a Lagrangian capable to produce
these roots via billiard walls needs more exotic fields than just
$p$-forms.
\newline

Our paper is organized as follows. The general framework of our
analysis is outlined in the first section: the form of the
searched for gravitational Lagrangians is recalled, together with
the list of their walls and the metric used to build the Cartan
matrix. In the next four sections, we deal with hyperbolic
algebras of rank 3 to 6. First, in $D=3$ spacetime dimensions, we
compute the 3-dimensional dilaton couplings needed to reproduce
the Dynkin diagram and check the status of the subdominant walls.
This is how we select the admissible algebras. Next, for each of
them, we determine which Lagrangian can be oxidized and we produce
it in the maximal oxidation dimension. The 18 hyperbolic algebras
of ranks 7 to 10 are reviewed in the last section; as explained
before, they are singled out for special treatment because 14 of
them are overextensions of finite dimensional simple Lie algebras
and the remaining 4 are dual to the overtextension $B_n^{\wedge\wedge}$
(with $n= 5,6,7,8$). We explicitly write down the $D_{max}=9$ Lagrangian
system obtained previously in \cite{MHBJ} the billiard of which is
the fundamental Weyl chamber of the algebra $CE_{10}$. Among the
four hyperbolic algebras of rank 10, $CE_{10}$ is special because,
unlike $E_{10}, BE_{10}$ and $DE_{10}$, it does not stem from
supergravities. Finally, we close our paper with some conclusions.

\section{General framework}

The billiard analysis refers to the dynamics, in the vicinity of a
spacelike singularity, of a gravitational model described by the
Lagrangian\footnote{Compared to the notations of \cite{DdBHS}, we
have put a factor of $2$ in the exponents of the dilaton
couplings; this way, a factor $1/2$ will be removed in front of the
dilatonic part of the $p$-form walls.}
\begin{eqnarray} {\cal L}_D &=& ^{(D)}R\star\unity -\sum_\alpha \star
d\phi^\alpha\wedge d\phi^\alpha - \nonumber \\ & \, &
-\frac{1}{2}\sum_p\,e^{2\lambda^{(p)} (\phi)}\star F^{(p+1)}\wedge
F^{(p+1)}, \quad D\geq 3\end{eqnarray} where $\lambda^{(p)} (\phi)
= \sum_\alpha \lambda^{(p)}_\alpha \phi^\alpha$ and $\star\unity =
\sqrt{\vert^{(D)}g\vert}\,dx^0\wedge...\wedge dx^{D-1}$. The
dilatons are denoted by $\phi^\alpha, (\alpha=1,...,N)$; their
kinetic terms are normalized with a weight 1 with respect to the
Ricci scalar. The Einstein metric has Lorentz signature
$(-,+,...,+)$; its determinant is $^{(D)}g$. The integer $p\geq 0$
labels the various $p$-forms $A^{(p)}$ present in the theory, with
field strengths $F^{(p+1)} = dA^{(p)}.$ If there are several
$p$-form gauge fields with the same form degree $p$, we will use
different letters $A^{(p)}, B^{(p)},..$ to distinguish
them.
\newline

The rules for computing the billiards have been given in details
in \cite{DHN, DHJN, DdBHS} to which we refer the reader. We here
recall the essential tools that are used throughout the paper.
\newline

\subsection{The walls}
The walls bounding the billiard have different origins:
some arise from the Einstein-Hilbert action and involve only the
scale factors $\beta^i,\,(i=1,...,d)$, introduced through the
Iwasawa decomposition of the space metric. They are

1) the symmetry walls \be w^S_{ij}(\beta) = \beta^j - \beta^i, \quad\quad
i<j,\ee and

2) the curvature walls
\be w^G_{i;jk}(\beta) = 2\beta^i +
\sum_{\ell\ne i,j,k}\beta^\ell,\quad\quad i\ne j, i\ne k, j\ne k.\ee The others
come from the energy densities of the $p$-forms; they depend on the scale
factors and the dilatons and are

3) the electric walls
 \be
w^{E(p)}_{i_1...i_p}(\beta,\phi) =
\beta^{i_1}+...+\beta^{i_p} + \sum_\alpha
\lambda^{(p)}_\alpha\,\phi^\alpha\quad \quad i_1<...<i_p\ee and

4) the magnetic walls \be w^{M(p)}_{i_1...i_{d-p-1}}(\beta,\phi) =
\beta^{i_1}+...+\beta^{i_{d-p-1}} - \sum_\alpha
\lambda^{(p)}_\alpha\,\phi^\alpha\quad \quad i_1<...<i_{d-p-1}.\ee

Notice that upon the change of $\phi^\alpha$ into $-\phi^\alpha$, the
electric walls of a $p$-form become the magnetic walls of its dual
$(d-p-1)$-form and vice versa.

The region of hyperbolic space where the
particle motion takes place is defined through the inequalties
$w_{ij}^S\geq 0$,
$w^G_{i;jk}\geq 0$, $w^{E(p)}_{i_1...i_p}\geq 0$ and
$w^{M(p)}_{i_1...i_{d-p-1}}\geq 0$; in fact, these inequalities
follow from a simpler subset, namely \be \beta^1\leq
\beta^2...\leq \beta^d,\quad w^G_{1;23}\geq 0,\quad
w^E_{1...p}\geq 0, \quad w^M_{1...(d-p-1)}\geq 0, \ee which may
still be redundant. The walls forming the minimal set needed to
define completely the billiard are called "dominant"; the others
are referred to as subdominant. More precisely, a wall is called
subdominant if it can be expressed as a linear combination with
positive coefficients of the dominant ones.

\subsection{The metric}
Given two walls $w(\beta, \phi) = w_i\,\beta^i +
w_\alpha\,\phi^\alpha = w_\mu\,\beta^\mu$ and $w'(\beta,\phi) =
w'_i\,\beta^i + w'_\alpha\,\phi^\alpha = w'_\mu\,\beta^\mu $ - the
$\beta^\mu (\mu=1,...,d,1+d,...,N+d)$ here denote scale factors
$\beta^i$ $(i=1,...,d)$ and dilatons,
$\beta^{\alpha+d}=\phi^\alpha $ - their scalar product is defined
as
\begin{eqnarray} (w\vert w') &=& G^{\mu\nu}\,w_\mu\,w'_\nu \nonumber\\ &=&
\sum_i\,(w_i\,w'_i) - \frac{1}{d-1}\,(\sum_i w_i)(\sum_j w'_j) +
\sum_\alpha (w_\alpha\,w'_\alpha).\label{metric}\end{eqnarray} The
metric $G^{\mu\nu}$ is the inverse of the Lorentzian metric
$G_{\mu\nu}$ defining the kinetic term of the scale factors and
dilatons; as shown in (\ref{metric}), it depends explicitly on the
spatial dimension $d$. Notice that a symmetry wall has a norm
squared equal to 2. Furthermore, the $p$-form electric wall
$w^E_{1...p}$ is orthogonal to all symmetry walls except one,
namely, $w^S_{p,p+1} = \beta^{p+1}-\beta^p$; the corresponding
scalar product is equal to $-1$.
\newline

Let $\{w_B = w_B(\beta,\phi),  B=1,...,r\}$ denote a set of dominant walls.
The enclosed billiard volume is finite if the scalar products are such that
the $r\times r$ matrix
\be A_{BC} = 2\,\frac{(w_B\vert w_C)}{(w_B\vert w_B)} \ee is the
generalized Cartan matrix of an hyperbolic Kac Moody algebra of rank $r$.
\newline

\section{Rank 3 Hyperbolic algebras}

\subsection{D=3}

In space dimension $d=2$, one has a single symmetry wall, namely

\be \alpha_1 =\beta^2- \beta^1\ee

and $N=r-d=3-2=1$ dilaton denoted as $\phi$. It
is obvious that only a $0$-form magnetic wall can be connected to the
symmetry wall, say

\be \alpha_2 = \beta^1-\lambda \phi.\ee

Let us show that the last dominant wall has to be an electric one
denoted by

\be\alpha_3=\lambda^{\prime} \phi.\label{ewa}\ee

Indeed, had one taken for dominant the magnetic wall
$\tilde\alpha_3=\beta^1-\lambda^{\prime}\phi$ instead of (\ref{ewa}), then,
its corresponding electric wall, which is precisely
$\alpha_3=\lambda^{\prime}\phi$, would be dominant too because of the
impossibility to write it as a linear combination with positive coefficients
of the other three $\alpha_1, \alpha_2$ and
$\tilde\alpha_3$.
\newline

Using the metric (\ref{metric}) adapted to
$d=2$, we build the matrix

\be A_{ij} = 2 { (\alpha_i \vert \alpha_j) \over (\alpha_i \vert \alpha_i)} \ee
and obtain

\be
 A = \left(
\begin{array}{ccc}
2 & -1 & 0 \\
-{2 \over \lambda^2} & 2 & -2 {\lambda^\prime \over \lambda} \\
0 & -2{ \lambda \over \lambda^\prime} & 2\\
 \end{array}
\right) \label{Amatrix}
\ee
which has to be identified with the generalized Cartan matrix of an
hyperbolic Kac Moody algebra of rank 3. Because $\phi$ can be
changed into $-\phi$,
$\lambda$ and $\lambda'$ can be chosen positive.

Since in such a matrix i) the non zero off-diagonal entries are
negative integers and ii) not any finite or affine Lie algebra of
rank 2 has an off-diagonal negative integer $<-4$, one immediately
infers from the expression of $A_{21}$ in (\ref{Amatrix}) that the
allowed values for $\lambda$ are \be \lambda \in
\{\sqrt{2},\, 1,\, \sqrt{2/3},\, 1/2\}.\ee Being a symmetry wall,
$\alpha_1$ has norm squared equal to $2$; $\alpha_2$ has norm
squared $\lambda^2\leq 2$, so that, if the Dynkin diagram has an
arrow between $\alpha_1$ and $\alpha_2$, this arrow must be
directed towards $\alpha_2$. Once the value of $\lambda$ has been
fixed, one needs to find $\lambda^\prime$ such that both
$2\lambda'/\lambda$ and $2\lambda/\lambda'$ are positive integers:
this leaves $\lambda = \lambda'/2, \lambda', 2\lambda'$. These
values are further constrained by the condition that the
subdominant walls $\tilde\alpha_2=\lambda \phi$ and
$\tilde\alpha_3=\beta^1 - \lambda^\prime \phi$, stay really behind
the others i.e., that there exist $k>0$ and $\ell\geq 0$ such that
\begin{eqnarray}\tilde\alpha_2 &=& k\alpha_3
\Longrightarrow \lambda/\lambda' = k\\ \tilde\alpha_3 &=&
\alpha_2+\ell\alpha_3
\Longrightarrow \lambda/\lambda' = \ell+1\end{eqnarray} which implies
$k=\ell+1\geq 1$. Hence, the subdominant conditions require
\be \lambda^\prime = \lambda\quad\mbox{or}\quad \lambda^\prime =
\lambda /2.\ee  Let us summarize the 8 different
possibilities for the pairs
$(\lambda, \lambda')$ that lead to Cartan matrices and draw the
corresponding Dynkin diagrams:
\newline

(i) for $\lambda = \sqrt{2}$ and $\lambda^\prime =
\sqrt{2}$, the Dynkin diagram describes the overextension
$A_1^{\wedge\wedge}$

\begin{center}
\scalebox{.5}{
\begin{picture}(180,60)
\put(-45,10){3-1}
\thicklines \multiput(10,10)(40,0){3}{\circle{10}}
\put(15,10){\line(1,0){30}}
\put(55,7.5){\line(1,0){30}} \put(55,12.5){\line(1,0){30}}
\end{picture}
}
\end{center}
and for $\lambda=\sqrt{2}$ and $\lambda^\prime= 1/\sqrt{2}$, the
Dynkin diagram corresponds to the twisted overextension \cite{HJ}
$A_2^{(2)\wedge}$
\begin{center}
\scalebox{.5}{
\begin{picture}(180,60)
\put(-45,10){3-2}
\thicklines \multiput(10,10)(40,0){3}{\circle{10}}
\put(15,10){\line(1,0){30}}
\put(55,8.75){\line(1,0){30}} \put(55,11.25){\line(1,0){30}}
\put(55,6.25){\line(1,0){30}} \put(55,13.75){\line(1,0){30}}
\put(65,0){\line(1,1){10}} \put(65,20){\line(1,-1){10}}
\end{picture}
}
\end{center}

(ii) $\lambda = 1$; the two possibilities are $\lambda^\prime = 1$ and
$\lambda^\prime= 1/2$. The Dynkin diagrams are respectively
\begin{center}
\scalebox{.5}{
\begin{picture}(180,60)
\put(-45,10){3-3}
\thicklines \multiput(10,10)(40,0){3}{\circle{10}}
\put(15,7.5){\line(1,0){30}} \put(15,12.5){\line(1,0){30}}
\put(25,0){\line(1,1){10}} \put(25,20){\line(1,-1){10}}
\put(55,7.5){\line(1,0){30}} \put(55,12.5){\line(1,0){30}}
\end{picture}
}
\end{center}
\begin{center}
\scalebox{.5}{
\begin{picture}(180,60)
\put(-45,10){3-4}
\thicklines \multiput(10,10)(40,0){3}{\circle{10}}
\put(15,12.5){\line(1,0){30}} \put(15,7.5){\line(1,0){30}}
\put(25,0){\line(1,1){10}} \put(25,20){\line(1,-1){10}}
\put(55,8.75){\line(1,0){30}} \put(55,11.25){\line(1,0){30}}
\put(55,6.25){\line(1,0){30}} \put(55,13.75){\line(1,0){30}}
\put(65,0){\line(1,1){10}} \put(65,20){\line(1,-1){10}}
\end{picture}
}
\end{center}

(iii) $\lambda = \sqrt{2/3}$; the two possibilities are
$\lambda^\prime = \sqrt{2/3}$ and $\lambda^\prime= 1/\sqrt{6}$
with Dynkin diagrams given by,
\begin{center}
\scalebox{.5}{
\begin{picture}(180,60)
\put(-45,10){3-5}
\thicklines \multiput(10,10)(40,0){3}{\circle{10}}
\put(15,7.5){\line(1,0){30}} \put(15,12.5){\line(1,0){30}}
\put(15,10){\line(1,0){30}}
\put(25,0){\line(1,1){10}} \put(25,20){\line(1,-1){10}}
\put(55,7.5){\line(1,0){30}} \put(55,12.5){\line(1,0){30}}
\end{picture}
}
\end{center}
\begin{center}
\scalebox{.5}{
\begin{picture}(180,60)
\put(-45,10){3-6}
\thicklines \multiput(10,10)(40,0){3}{\circle{10}}
\put(15,12.5){\line(1,0){30}} \put(15,7.5){\line(1,0){30}}
\put(15,10){\line(1,0){30}}
\put(25,0){\line(1,1){10}} \put(25,20){\line(1,-1){10}}
\put(55,8.75){\line(1,0){30}} \put(55,11.25){\line(1,0){30}}
\put(55,6.25){\line(1,0){30}} \put(55,13.75){\line(1,0){30}}
\put(65,0){\line(1,1){10}} \put(65,20){\line(1,-1){10}}
\end{picture}
}
\end{center}

(iv) $\lambda = 1/2$; the two possibilities are $\lambda^\prime = 1/2$
and
$\lambda^\prime= 1/4$. The Dynkin diagrams are respectively,
\begin{center}
\scalebox{.5}{
\begin{picture}(180,60)
\put(-45,10){3-7}
\thicklines \multiput(10,10)(40,0){3}{\circle{10}}
\put(15,8.75){\line(1,0){30}} \put(15,11.25){\line(1,0){30}}
\put(15,6.25){\line(1,0){30}} \put(15,13.75){\line(1,0){30}}
\put(25,0){\line(1,1){10}} \put(25,20){\line(1,-1){10}}
\put(55,7.5){\line(1,0){30}} \put(55,12.5){\line(1,0){30}}
\end{picture}
}
\end{center}
\begin{center}
\scalebox{.5}{
\begin{picture}(180,60)
\put(-45,10){3-8}
\thicklines \multiput(10,10)(40,0){3}{\circle{10}}
\put(15,8.75){\line(1,0){30}} \put(15,11.25){\line(1,0){30}}
\put(15,6.25){\line(1,0){30}} \put(15,13.75){\line(1,0){30}}
\put(25,0){\line(1,1){10}} \put(25,20){\line(1,-1){10}}
\put(55,8.75){\line(1,0){30}} \put(55,11.25){\line(1,0){30}}
\put(55,6.25){\line(1,0){30}} \put(55,13.75){\line(1,0){30}}
\put(65,0){\line(1,1){10}} \put(65,20){\line(1,-1){10}}
\end{picture}
}
\end{center}

\textbf{Comments}\newline

1) When $\lambda^\prime=\lambda$, $\alpha_2$ and $\alpha_3$ have
to be assigned to a single scalar field; when
$\lambda^\prime\ne\lambda$, two scalars are needed in the 3-dimensional
Lagrangian.

2) The algebra $(3-8)$ is missing in table 2 of reference
\cite{S}. The subalgebra obtained when removing the first or the
last root is the affine $A_2^{(2)}$; the removal of the middle
root gives $A_1\times A_1$ so that this algebra satisfies indeed
the criterion of hyperbolicity.

3) Remark that none of the 8 algebras above is strictly hyperbolic
\footnote{A strictly hyperbolic algebra is such that upon removal of a
simple root, only finite Lie algebras are left
behind.}. The latter are listed in table 1 of \cite{S}.

\subsection{D=4}

The 4-dimensional Lagrangian will have no dilaton in it since $N=r-d=0$;
hence, if such a Lagrangian exists, it cannot stem from an higher dimensional
parent and $D_{max}=4$. When looking at the algebras of rank 3 selected
above, one sees that only
$(3-1)$ and
$(3-2)$ have an A-chain of length $k=2$ and allow, a priori, a
second symmetry wall. One starts with \be
\alpha_1=\beta^3-\beta^2\quad\mbox{and}\quad
\alpha_2=\beta^2-\beta^1.\ee

The third root may only contain $\beta^1$ and can be associated to
\begin{enumerate}
\item{the curvature wall
$\alpha_3 = 2\beta^1$ in the case of 4-dimensional pure gravity.}
The Dynkin diagram bears number $(3-1)$ above and is the
overextension $A_1^{\wedge\wedge}$.
\item{the electric/magnetic wall of a $1$-form: $\alpha_3 = \beta^1$.}
This case leads to diagram $(3-2)$ which belongs to the twisted overextension
$A_2^{(2)\wedge}$.
\end{enumerate}

One sees immediately that the regions of hyperbolic space
delimited by both sets of walls coincide; the difference is
entirely due to the normalization of the third wall which is thus
responsible for the emergence of two distinct Cartan matrices.

\section{Rank 4 Hyperbolic algebras}
\subsection{D=3}

In order to reproduce through walls the four roots of such an
algebra, besides the scale factors $\beta^1$ and $\beta^2$, one
needs $N=2$ dilatons; they will be denoted as
$\phi^1=\phi,\,\,\phi^2 = \varphi$. There is one symmetry wall,
i.e. $\alpha_1 = \beta^2-\beta^1$ and, a priori, two choices can
be made for the next three dominant walls: either (i) one takes
one magnetic wall and two electric ones or (ii) one takes one
electric wall and two magnetic ones. We will start with case (i)
and show later how case (ii) is eliminated on account of the
subdominant conditions.

\subsubsection{One magnetic wall and two electric ones}

The dominant walls are thus the symmetry wall \be \alpha_1 =
\beta^2-\beta^1,\ee the magnetic wall, written as\footnote{This
ansatz represents no loss of generality because starting from the
more general expression $\alpha_2 = \beta^1 - \lambda \phi +\mu
\varphi$, one can redefine the dilatons via an
orthogonal transformation - leaving the dilaton
Lagrangian invariant - to get the simpler expression used above.} \be
\alpha_2 = \beta^1 - \lambda \phi \ee and the two electric ones
\be \alpha_3=\lambda'\phi - \mu' \varphi \ee respectively \be
\alpha_4=\lambda'' \phi +\mu'' \varphi.\ee As before, the signs
have already been distributed to account for the negative signs of
the off-diagonal Cartan matrix elements when allowing the
parameters to be either all $\geq 0$ or all $\leq 0$; that they
can further be chosen positive is due the possibility to change
$\phi^\alpha$ into $-\phi^\alpha$.
The general structure of the Dynkin diagram is therefore the
following

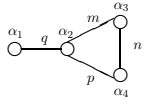
\begin{center}
\scalebox{.5}{
\begin{picture}(180,60)
\thicklines \multiput(10,10)(40,0){2}{\circle{10}}
\put(90,30){\circle{10}} \put(90,-10){\circle{10}}
\put(90,-5){\line(0,1){30}} \put(15,10){\line(1,0){30}}
\put(50,15){\line(2,1){35}} \put(50,5){\line(2,-1){35}}
\put(30,15){$q$} \put(100,10){$n$} \put(65,30){$m$}
\put(65,-15){$p$} \put(5,20){$\alpha_1$} \put(43,20){$\alpha_2$}
 \put(85,-25){$\alpha_4$} \put(85,40){$\alpha_3$}
\end{picture}
}
\end{center}\vspace{1cm}
where we have not drawn the arrows and $q, \ m, \ n, \ p$ are
integers which count the number of lines joining two vertices.
\newline

What are the possible values that can be assigned to $q$, $m$, $n$
and $p$? Since this diagram has to become the Dynkin diagram of an
hyperbolic algebra, the maximal value of each of these integers is
$3$, because there is no finite or affine algebra of rank 3 with
off-diagonal Cartan matrix elements smaller than $-3$. Another
point is that if there were an arrow between $\alpha_1$ and
$\alpha_2$ it necessarily points towards $\alpha_2$: one has
indeed $(\alpha_1,\alpha_2)=-1$, $(\alpha_1,\alpha_1)=2$ (it is a
symmetry wall), $(\alpha_2,\alpha_2) = \lambda^2$ and $A_{21}
=-2/\lambda^2$ can only be $-1,-2$ or $-3$. We may also state that
if $A_{ij}$ is neither $0$ nor $-1$ then $A_{ji} = -1$, because
this is a common property of all finite or affine algebras of rank
3. Taking all these restrictions into account, one has to consider
three different situations characterized respectively by (i) $m$,
$n$, $p$ are all different from zero, (ii) $n=0$ and $m$, $p$ are
not zero (iii) $p=0$ and $n$, $m$ are not zero.
\newline

(i) If $m$, $n$ and $p$ are all non zero, then they must all be equal to 1
because, upon removal of the root $\alpha_1$, one obtains a triangular
diagram; now, in the set of the finite or
affine algebras, there is only one such triangular Dynkin diagram and it is
simply laced. That leaves, a priori, three cases labelled by the values
$q=1,2,3$. The corresponding dilaton couplings are

\be
\lambda =\sqrt{ {2 \over q}};\quad
\lambda' = {1 \over \sqrt{2q}} ;\quad \mu' = \sqrt{ {3 \over 2 q}};\quad
\lambda''= {1 \over \sqrt{2q}} ;\quad  \mu'' = \sqrt{ {3 \over 2
q}}.\label{i} \ee

The Dynkin diagrams corresponding to $q=1$, $2$ and $3$ are
respectively,

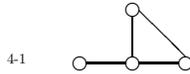
\begin{center}
\scalebox{.5}{
\begin{picture}(180,60)
\put(-45,10){4-1}
\thicklines \multiput(10,10)(40,0){3}{\circle{10}}
\multiput(15,10)(40,0){2}{\line(1,0){30}}
\put(50,50){\circle{10}} \put(50,15){\line(0,1){30}}
\put(55,50){\line(1,-1){35}}
\end{picture}
}
\end{center}
which is the overextension $A_2^{\wedge\wedge}$ and
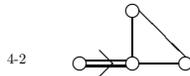
\begin{center}
\scalebox{.5}{
\begin{picture}(180,60)
\put(-45,10){4-2}
\thicklines \put(15,8){\line(1,0){30}} \put(15,12){\line(1,0){30}}
\multiput(10,10)(40,0){3}{\circle{10}} \put(25,0){\line(1,1){10}}
\put(25,20){\line(1,-1){10}} \put(55,10){\line(1,0){30}}
\put(50,50){\circle{10}} \put(50,15){\line(0,1){30}}
\put(55,50){\line(1,-1){35}}
\end{picture}
}
\end{center}

\begin{center}
\scalebox{.5}
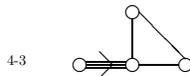
{
\begin{picture}(180,60)
\put(-45,10){4-3}
 \thicklines \put(15,7.5){\line(1,0){30}}
\put(15,12.5){\line(1,0){30}}
\multiput(10,10)(40,0){3}{\circle{10}} \put(25,0){\line(1,1){10}}
\put(25,20){\line(1,-1){10}}
\multiput(15,10)(40,0){2}{\line(1,0){30}}
\put(50,50){\circle{10}} \put(50,15){\line(0,1){30}}
\put(55,50){\line(1,-1){35}}
\end{picture}
}
\end{center}

The subdominant conditions are satisfied in all cases; let us show
this explicitly. With the couplings in (\ref{i}), the dominant
walls other than the symmetry wall read \be \alpha_2 = \beta^1 -
2\,\frac{\phi}{\sqrt{2q}}\quad,\quad \alpha_3 =
\frac{\phi}{\sqrt{2q}} -
\varphi\sqrt{\frac{3}{2q}}\quad,\quad\alpha_4 =
\frac{\phi}{\sqrt{2q}} + \varphi\sqrt{\frac{3}{2q}}.\ee The
corresponding subdominant ones are \be \tilde\alpha_2 =
 2\,\frac{\phi}{\sqrt{2q}}\quad,\quad
\tilde\alpha_3 = \beta^1-\frac{\phi}{\sqrt{2q}} +
\varphi\sqrt{\frac{3}{2q}}
\quad,\quad \tilde\alpha_4=\beta^1-\frac{\phi}{\sqrt{2q}} -
\varphi\sqrt{\frac{3}{2q}}\ee and they obey \be \tilde\alpha_2 =
\alpha_3+\alpha_4\quad,\quad \tilde\alpha_3 =
\alpha_2+\alpha_4\quad,\quad\tilde\alpha_4 = \alpha_2+\alpha_3.\ee
\newline

(ii) if $n=0$ and $m$, $p$ are not zero, the structure of the
Dynkin diagram is the following
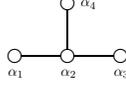
\begin{center}
\scalebox{.5}{
\begin{picture}(180,60)
\thicklines \multiput(10,10)(40,0){3}{\circle{10}}
\put(15,10){\line(1,0){30}}
\put(55,10){\line(1,0){30}}
\put(50,50){\circle{10}}
\put(50,15){\line(0,1){30}}
\put(5,-5){$\alpha_1$} \put(45,-5){$\alpha_2$}
 \put(60,47.5){$\alpha_4$} \put(85,-5){$\alpha_3$}
\end{picture}
}
\end{center}

Comparison with the similar graphs of \cite{S} impose (i) $m=p=2$,
(ii) $q=1$ or $q=2$ and (iii) an arrow pointing from
$\alpha_2$ to $\alpha_3$ and another arrow from $\alpha_2$ to $\alpha_4$.
Accordingly, the dilaton couplings producing them are given by

\be
\lambda = \sqrt{ {2 \over q}};\quad
\lambda' =  {1 \over \sqrt{2q}} ;\quad \mu' = {1 \over \sqrt{2q}} ;\quad
\lambda''= {1 \over \sqrt{2q}} ;\quad \mu'' = {1 \over \sqrt{2q}}.
\label{ii}
\ee

The Dynkin diagrams corresponding to $q=1$ and $2$ are respectively,


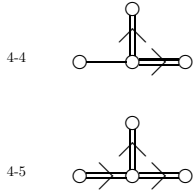
\begin{center}
\scalebox{.5}{
\begin{picture}(180,60)
\put(-45,10){4-4}
\thicklines \multiput(10,10)(40,0){3}{\circle{10}}
\put(15,10){\line(1,0){30}}
\put(55,8){\line(1,0){30}} \put(55,12){\line(1,0){30}}
\put(50,50){\circle{10}}
\put(48,15){\line(0,1){30}} \put(52,15){\line(0,1){30}}
\put(65,0){\line(1,1){10}} \put(65,20){\line(1,-1){10}}
\put(40,25){\line(1,1){10}} \put(50,35){\line(1,-1){10}}
\end{picture}
}
\end{center}


\begin{center}
\scalebox{.5}{
\begin{picture}(180,60)
\put(-45,10){4-5}
\thicklines \put(15,8){\line(1,0){30}} \put(15,12){\line(1,0){30}}
\multiput(10,10)(40,0){3}{\circle{10}} \put(25,0){\line(1,1){10}}
\put(25,20){\line(1,-1){10}} \put(55,12){\line(1,0){30}}
\put(55,8){\line(1,0){30}}
\put(50,50){\circle{10}}
\put(48,15){\line(0,1){30}} \put(52,15){\line(0,1){30}}
\put(65,0){\line(1,1){10}} \put(65,20){\line(1,-1){10}}
\put(40,25){\line(1,1){10}} \put(50,35){\line(1,-1){10}}
\end{picture}
}
\end{center}
Again, the subdominant conditions are fulfilled: indeed, one gets
$\tilde\alpha_2=\alpha_3+\alpha_4$, $\tilde\alpha_3=\alpha_2+\alpha_4$,
$\tilde\alpha_4=\alpha_2+\alpha_3$.
\newline

(iii) if $p=0$ and $n$, $m$ are not zero, the structure of the
Dynkin diagram is the following
\begin{center}
\scalebox{.5}{
\begin{picture}(180,60)
\thicklines \multiput(10,10)(40,0){4}{\circle{10}}
\multiput(15,10)(40,0){3}{\line(1,0){30}}
\put(5,-5){$\alpha_1$} \put(45,-5){$\alpha_2$}
 \put(125,-5){$\alpha_4$} \put(85,-5){$\alpha_3$}
\end{picture}
}
\end{center}

The dominant walls now simplify as
\be\alpha_1 =
\beta^2-\beta^1 \quad,\quad \alpha_2 = \beta^1 - \lambda
\phi
\quad,\quad \alpha_3 =\lambda'\phi - \mu'
\varphi\quad,\quad \alpha_4 = \mu'' \varphi.
\ee
 We want the
corresponding magnetic and electric walls to be effectively
subdominant: this is indeed satisfied when 1) $\lambda$ and
therefore $\lambda'$ are positive; 2) $\lambda' / \lambda \leq 1$
(that is $\lambda' / \lambda = 1$ or $\lambda' / \lambda =1/2$)
and 3) $\mu' / \mu'' \geq \lambda' / \lambda$. Accordingly, the
remaining possibilities for $\lambda$, $\lambda'$ and $\mu'$ are,
a priori, those given in the following table:

\begin{center}
\begin{tabular}{|c|c|c|c|}
\hline & $\lambda$ & $\lambda'$  & $ \mu'$ \\
\hline 1 &  $\sqrt{2}$ & $\sqrt{2}$ & $\sqrt{2}$  \\
\hline 2.a &  $\sqrt{2}$ & ${1 / \sqrt{2}}$ & $\sqrt{3 / 2}$  \\
\hline 2.b &  $\sqrt{2}$ & ${1 / \sqrt{2}}$ & ${1 / \sqrt{2}}$  \\
\hline 2.c &  $\sqrt{2}$ & ${1 / \sqrt{2}}$ & ${1 / \sqrt{6}}$  \\
\hline 3 &1 & 1& 1  \\
\hline 4.a  &1 & ${1/ 2}$ & ${  \sqrt{3}/ 2}$  \\
\hline  4.b &1 & ${1/ 2}$ & ${ 1 / 2}$  \\
\hline 4.c  &1 & ${1/2}$ & ${ 1 /\sqrt{12}}$  \\
\hline 5 & $ \sqrt{ {2 / 3}}$ &  $ \sqrt{ {2 / 3}}$ & $ \sqrt{ {2 /
3}}$  \\
\hline 6.a &  $ \sqrt{ {2 / 3}}$ &  $  {1 / \sqrt{6}}$ &
${1 / \sqrt{2}}$ \\
\hline 6.b&  $ \sqrt{ 2 / 3}$ &  $  {1 / \sqrt{6}}$ & ${1
/ \sqrt{6}}$ \\
\hline 6.c&  $ \sqrt{ {2 / 3}}$ & $  {1 / \sqrt{6}}$ & ${1
/ \sqrt{18}}$\\
\hline
\end{tabular}
\end{center}

The different values for $\mu'$ correspond to distinct admissible
values for $A_{32}$. Finally, for the values of $\mu''$, we again
meet two cases depending on which of $A_{34}$ or $A_{43}$ is equal
to $-1$. In each case, one has still to check the subdominant
conditions.\newline

1. The two possibilities lead to a Cartan matrix: either $\mu'' =
2\sqrt{2}$ or $\mu'' = \sqrt{2}$. The former case is ruled out
because the subdominant conditions cannot be satisfied. The Dynkin
diagram of the remaining case describes the twisted overextension
$D_3^{(2)\wedge}$

\begin{center}
\scalebox{.5}{
\begin{picture}(180,60)
\put(-45,10){4-6} \thicklines
\multiput(10,10)(40,0){4}{\circle{10}} \put(15,10){\line(1,0){30}}
\put(55,12.5){\line(1,0){30}} \put(55,7.5){\line(1,0){30}}
\put(65,10){\line(1,1){10}} \put(65,10){\line(1,-1){10}}
\put(95,12.5){\line(1,0){30}} \put(95,7.5){\line(1,0){30}}
\put(105,0){\line(1,1){10}} \put(105,20){\line(1,-1){10}}
\end{picture}
}
\end{center}

2. a. Either $\mu'' = \sqrt{2/3}$ or $\mu'' =
\sqrt{6}$; both lead to hyperbolic algebras which correspond
respectively to the overextension $G_2^{\wedge\wedge}$

\begin{center}
\scalebox{.5}{
\begin{picture}(180,60)
\put(-45,10){4-7 } \thicklines
\multiput(10,10)(40,0){4}{\circle{10}}
\multiput(15,10)(40,0){2}{\line(1,0){30}}
\put(95,12.5){\line(1,0){30}} \put(95,7.5){\line(1,0){30}}
\put(95,10){\line(1,0){30}}
\put(105,0){\line(1,1){10}} \put(105,20){\line(1,-1){10}}
\end{picture}
}
\end{center}
and to the twisted overextension $D_4^{(3)\wedge}$
\begin{center}
\scalebox{.5}{
\begin{picture}(180,60)
\put(-45,10){4-8 } \thicklines
\multiput(10,10)(40,0){4}{\circle{10}}
\multiput(15,10)(40,0){2}{\line(1,0){30}}
\put(95,12.5){\line(1,0){30}} \put(95,10){\line(1,0){30}}
\put(95,7.5){\line(1,0){30}} \put(105,10){\line(1,1){10}}
\put(105,10){\line(1,-1){10}}
\end{picture}
}
\end{center}

2. b. Either $\mu'' = \sqrt{1/2}$ or $\mu'' =
\sqrt{2}$; the Dynkin diagrams correspond respectively to
the twisted overextension $A_4^{(2)\wedge}$

\begin{center}
\scalebox{.5}{
\begin{picture}(180,60)
\put(-45,10){4-9} \thicklines
\multiput(10,10)(40,0){4}{\circle{10}} \put(15,10){\line(1,0){30}}
\put(55,12.5){\line(1,0){30}} \put(55,7.5){\line(1,0){30}}
\put(65,0){\line(1,1){10}} \put(65,20){\line(1,-1){10}}
\put(95,12.5){\line(1,0){30}} \put(95,7.5){\line(1,0){30}}
\put(105,0){\line(1,1){10}} \put(105,20){\line(1,-1){10}}
\end{picture}
}
\end{center}

and to the overextension $C_2^{\wedge\wedge}$
\begin{center}
\scalebox{.5}{
\begin{picture}(180,60)
\put(-45,10){4-10}\thicklines
\multiput(10,10)(40,0){4}{\circle{10}} \put(15,10){\line(1,0){30}}
\put(55,12.5){\line(1,0){30}} \put(55,7.5){\line(1,0){30}}
\put(65,0){\line(1,1){10}} \put(65,20){\line(1,-1){10}}
\put(95,12.5){\line(1,0){30}} \put(95,7.5){\line(1,0){30}}
\put(105,10){\line(1,1){10}} \put(105,10){\line(1,-1){10}}
\end{picture}
}
\end{center}

2. c. Only the value $\mu''= 2/\sqrt{6}$ is
compatible with the subdominant conditions. The corresponding algebra is
given by

\begin{center}
\scalebox{.5}{
\begin{picture}(180,60)
\put(-45,10){4-11}
\thicklines \multiput(10,10)(40,0){4}{\circle{10}}
\multiput(15,10)(40,0){3}{\line(1,0){30}}
\put(55,12.5){\line(1,0){30}} \put(55,7.5){\line(1,0){30}}
\put(65,0){\line(1,1){10}} \put(65,20){\line(1,-1){10}}
\end{picture}
}
\end{center}

3. Here again, only the value $\mu''=1$ can be retained on account of the
subdominant conditions. This leads to

\begin{center}
\scalebox{.5}{
\begin{picture}(180,60)
\put(-45,10){4-12}
\thicklines \multiput(10,10)(40,0){4}{\circle{10}}
\put(15,12){\line(1,0){30}} \put(15,8){\line(1,0){30}}
\put(25,0){\line(1,1){10}} \put(25,20){\line(1,-1){10}}
\put(55,12.5){\line(1,0){30}} \put(55,7.5){\line(1,0){30}}
\put(65,10){\line(1,1){10}} \put(65,10){\line(1,-1){10}}
\put(95,12.5){\line(1,0){30}} \put(95,7.5){\line(1,0){30}}
\put(105,0){\line(1,1){10}} \put(105,20){\line(1,-1){10}}
\end{picture}
}
\end{center}

4. a. Either $\mu'' = \sqrt{3}$ or $\mu'' = 1/
\sqrt{3}$; both values are admissible. They lead to

\begin{center}
\scalebox{.5}{
\begin{picture}(180,60)
\put(-45,10){4-13}
\thicklines \multiput(10,10)(40,0){4}{\circle{10}}
\put(15,12){\line(1,0){30}} \put(15,8){\line(1,0){30}}
\put(25,0){\line(1,1){10}} \put(25,20){\line(1,-1){10}}
\put(55,10){\line(1,0){30}}
\put(95,12.5){\line(1,0){30}} \put(95,7.5){\line(1,0){30}}
\put(95,10){\line(1,0){30}} \put(105,10){\line(1,1){10}}
\put(105,10){\line(1,-1){10}}
\end{picture}
}
\end{center}

\begin{center}
\scalebox{.5}{
\begin{picture}(180,60)
\put(-45,10){4-14}
\thicklines \multiput(10,10)(40,0){4}{\circle{10}}
\put(15,12){\line(1,0){30}} \put(15,8){\line(1,0){30}}
\put(25,0){\line(1,1){10}} \put(25,20){\line(1,-1){10}}
\put(55,10){\line(1,0){30}}
\put(95,12.5){\line(1,0){30}} \put(95,7.5){\line(1,0){30}}
\put(95,10){\line(1,0){30}} \put(105,0){\line(1,1){10}}
\put(105,20){\line(1,-1){10}}
\end{picture}
}
\end{center}

4. b. Either $\mu'' = 1/2$ or $\mu'' = 1$; both values are allowed and
they give respectively

\begin{center}
\scalebox{.5}{
\begin{picture}(180,60)
\put(-45,10){4-15}
\thicklines \multiput(10,10)(40,0){4}{\circle{10}}
\put(15,12){\line(1,0){30}} \put(15,8){\line(1,0){30}}
\put(25,0){\line(1,1){10}} \put(25,20){\line(1,-1){10}}
\put(55,12.5){\line(1,0){30}} \put(55,7.5){\line(1,0){30}}
\put(65,0){\line(1,1){10}} \put(65,20){\line(1,-1){10}}
\put(95,12.5){\line(1,0){30}} \put(95,7.5){\line(1,0){30}}
\put(105,0){\line(1,1){10}} \put(105,20){\line(1,-1){10}}
\end{picture}
}
\end{center}

\begin{center}
\scalebox{.5}{
\begin{picture}(180,60)
\put(-45,10){4-16}
\thicklines \multiput(10,10)(40,0){4}{\circle{10}}
\put(15,12.5){\line(1,0){30}} \put(15,7.5){\line(1,0){30}}
\put(25,0){\line(1,1){10}} \put(25,20){\line(1,-1){10}}
\put(55,12.5){\line(1,0){30}} \put(55,7.5){\line(1,0){30}}
\put(65,0){\line(1,1){10}} \put(65,20){\line(1,-1){10}}
\put(95,12.5){\line(1,0){30}} \put(95,7.5){\line(1,0){30}}
\put(105,10){\line(1,1){10}} \put(105,10){\line(1,-1){10}}
\end{picture}
}
\end{center}

4. c. Does not correspond to any hyperbolic algebra. \newline

5. Only the value $\mu''= \sqrt{2/3}$ is compatible with the
subdominant conditions but again there is no corresponding
hyperbolic algebra.
\newline

6. a. Only the first of the 2 values $\mu''=\sqrt{2}$ and
$\mu''=\sqrt{2}/3$ leads to an hyperbolic algebra, which is

\begin{center}
\scalebox{.5}{
\begin{picture}(180,60)
\put(-45,10){4-17}
\thicklines \multiput(10,10)(40,0){4}{\circle{10}}
\put(15,10){\line(1,0){30}} \put(15.5,12.5){\line(1,0){30}}
\put(15,7.5){\line(1,0){30}} \put(25,0){\line(1,1){10}}
\put(25,20){\line(1,-1){10}}
\put(55,10){\line(1,0){30}}
\put(95,12.5){\line(1,0){30}} \put(95,7.5){\line(1,0){30}}
\put(95,10){\line(1,0){30}} \put(105,10){\line(1,1){10}}
\put(105,10){\line(1,-1){10}}
\end{picture}
}
\end{center}

6. b. and 6. c. do not give an hyperbolic algebra.
\newline

\subsubsection{One electric wall and two magnetic ones \label{EMM}}
This case can be eliminated on account of the subdominant
conditions. Indeed, without loss of generality, one may choose the
parametrization of the dominant walls such that the electric wall
takes a simple form, i.e., such that
\begin{eqnarray}
\alpha_1 &=& \beta^2 - \beta^1 \nonumber\\
\alpha_2&=&\beta^1 -\lambda\,\phi -\mu\,\varphi \nonumber\\
\alpha_3 &=&\beta^1 -\lambda'\,\phi +\mu'\,\varphi \nonumber\\
\alpha_4 &=& \lambda''\,\phi.\label{428}\end{eqnarray} Being
assumed subdominant, the electric walls associated to $\alpha_2$
and $\alpha_3$, namely $\tilde \alpha_2 = \lambda\,\phi +
\mu\,\varphi$ and $\tilde \alpha_3 = \lambda'\,\phi
-\mu'\,\varphi$ need be proportional to $\alpha_4$; this happens
only when $\mu=\mu'=0$,  but then (\ref{428}) does no longer
define a rank four root system.

\subsection{$D>3$}

Our aim is now to determine which of the 17 algebras selected in
the previous section admit Lagrangians in higher spacetime
dimensions and to provide the maximal oxidation dimension and the
$p$-forms content with its characteristic features. By considering
each Dynkin diagram and looking at the length of the A-chain
starting from the symmetry root $\alpha_1$, we establish the
following "empirical" oxidation rule:  if the A-chain has length
$k$ one can oxidize the spatial dimension up to (i) $d=k+1$ if the
norm squared of the next connected root is smaller than $2$ and up
(ii) to $d=k$ if the norm squared of the next connected root is
greater than $2$. In particular, the subdominant conditions are
always satisfied. Explicitly,
\newline
\begin{enumerate}
\item{Diagram $(4-1)$} is the overextension
$A_2^{\wedge\wedge}$. We know from \cite{DdBHS} that it
corresponds to pure gravity in
$D_{max}=5$
\item{Diagrams $(4-2)$ and $(4-3)$} have an A-chain of length 1; the
3-dimensional theory cannot be oxidized.
\item{Diagram $(4-4)$} : $D_{max}=4$. The walls are given by
\begin{eqnarray}\alpha_1 &=&
\beta^3-\beta^2,
\quad\quad\quad\quad
\alpha_2=
\beta^2-\beta^1,\\ \alpha_3 &=& \beta^1 - \phi/ \sqrt{2},\quad\quad
\alpha_4=\beta^1 + \phi/ \sqrt{2} .\end{eqnarray} The last two are the
electric and magnetic dominant walls of a one-form coupled to the dilaton.
One sees immediately that $\tilde\alpha_3 = \alpha_4$ and
$\tilde\alpha_4=\alpha_3$.
\item{Diagram $(4-5)$} : the 3-dimensional Lagrangian has
no parent in
$D>3$.
\item{Diagram $(4-6)$} is the twisted overextension $D_3^{(2)\wedge}$.
The 3-D Lagrangian cannot be oxidized the reason being that
$\Vert\alpha_3\Vert^2>2$.
\item{Diagram $(4-7)$} is the overextension $G_2^{\wedge \wedge}$. We
know from \cite{DdBHS} that the theory can be oxidized up to $D_{max}= 5$
where the Lagrangian is that of the Einstein-Maxwell system.
\item{Diagram $(4-8)$} describes $D_4^{(3)\wedge}$. The A-chain has length
$k=3$ and the next connected root is longer than $\sqrt{2}$. The maximal
oxidation dimension is
$D_{max}=4$ and the dominant walls are given by
\begin{eqnarray}
\alpha_1 &=&
\beta^3-\beta^2,\quad\quad\quad\,\,\,\,\, \alpha_2= \beta^2-\beta^1,\\
\alpha_3 &=&
\beta^1 - \sqrt{3/2} \phi, \quad\quad \alpha_4= \sqrt{6} \phi
\end{eqnarray} The root $\alpha_3$ is the electric wall of a
$1$-form,
$\alpha_4$ is the electric wall of a $0$-form. One easily checks that
the subdominant magnetic walls satisfy \begin{eqnarray}\tilde\alpha_3 &=&
\beta^1+
\sqrt{3/2}
\phi\,\, =\,\, \alpha_3+\alpha_4 \\
\tilde\alpha_4 &=&
\beta^1+\beta^2-\sqrt{6} \phi\,\, =\,\, 2\alpha_3+\alpha_2.\end{eqnarray}
\item{Diagram $(4-9)$} represents $A_4^{(2)\wedge}$. $D_{max}=4$.
Its billiard realization requires
\begin{eqnarray} \alpha_1 &=&
\beta^3-\beta^2,\quad\quad\quad\quad \alpha_2= \beta^2-\beta^1,\\
\alpha_3 &=&
\beta^1 - \sqrt{1/2} \phi,\quad\quad\alpha_4 = \sqrt{1/2} \phi.
\end{eqnarray} The last two are again the electric walls of a $1$-form and
a zero-form; only the dilaton couplings differ from the previous
ones. The subdominant conditions are fulfilled: indeed, one obtains
$\tilde\alpha_3=\alpha_3+2\alpha_4$ and
$\tilde\alpha_4=2\alpha_3+\alpha_4+\alpha_2$.
\item{Diagram $(4-10)$} is the overextension $C_2^{\wedge\wedge}$. We
know from \cite{DdBHS} that the theory can be oxidized up to $D_{max}=4$.
\item{Diagram $(4-11)$} has $D_{max}=4$ and
\begin{eqnarray}\alpha_1 &=&
\beta^3-\beta^2,\quad\quad\quad\quad \alpha_2= \beta^2-\beta^1,\\
\alpha_3 &=& \beta^1 - \sqrt{1/6} \phi, \quad\quad\alpha_4=
\sqrt{2/3} \phi .\end{eqnarray} It has the same form content as
$(4-8)$ and $(4-9)$ but the dilaton couplings are still different.
Again, the subdominant conditions are
fulfilled: $\tilde\alpha_3=\alpha_3+\alpha_4$ and
$\tilde\alpha_4=2\alpha_3+\alpha_2$.
\item{Diagrams $(4-12)$ to $(4-17)$} : their 3-D Lagrangians cannot be
oxidized because there is a unique root of norm squared equal to 2.
\end{enumerate}

\textbf{Comments}\newline

a) The subdominant conditions are indeed always fulfilled in $D>3$ and
only positive integer coefficients enter the linear combinations.

b) In case $(4-4)$, $\alpha_3$ and $\alpha_4$ are the electric and
magnetic walls of the same one-form. In the other cases, they are
respectively assigned to a single one-form and a single zero-form.
The root multiplicity being one, there is no room for various
$p$-forms with identical couplings.

\section{Rank 5 Hyperbolic algebras}

\subsection{$D=3$}

The 3-dimensional Lagrangians need $N=r-d=3$ dilatons
($\phi^1=\phi, \phi^2=\varphi, \phi^3=\psi$); there are two scale
factors and one symmetry wall $\alpha_1= \beta^2-\beta^1$. In
order to reproduce the other four simple roots of the algebra in
terms of dominant walls, one has a priori three different cases to
consider: indeed, the set of dominant walls can comprise (i) one
magnetic wall and three electric ones, (ii) two electric walls and
two magnetic ones and (iii) one electric wall and three magnetic
ones. Only the first possibility will survive because as soon as
the set of dominant walls contains more than one magnetic wall,
one can show that the corresponding electric walls cannot fulfill
the subdominant conditions. Although the proof is a
straightforward generalization of the one given in subsection
(\ref{EMM}), we will provide it at the end of this section.

\subsubsection{One magnetic wall and three electric ones}

As in the previous sections, we use the
freedom to redefine dilatons through an orthogonal transformation and
choose the parametrization of the dominant walls such that
\begin{eqnarray}\alpha_1 &=& \beta^2-\beta^1 \\
\alpha_2 &=& \beta^1 - \lambda \phi\\\alpha_3 &=&\lambda'\phi - \mu'
\varphi\\ \alpha_4 &=&\lambda'' \phi + \mu'' \varphi - \nu''
\psi\\ \alpha_5 &=& \lambda''' \phi + \mu''' \varphi + \nu'''
\psi.\end{eqnarray}

One sees immediately that the symmetry root $\alpha_1$ is only linked to
the magnetic root $\alpha_2$ while
$\alpha_2$ can further be connected to one, two or three roots.  According
to
\cite{S}, five different structures for the Dynkin diagrams can be
encountered; we classify them below according to the total number of roots
connected to
$\alpha_2$; this number is 4 in case A, 3 in cases B and C, 2 in cases D
and E.
\newline

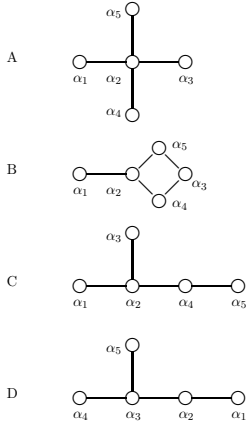
\begin{center} \scalebox{.5}{
\begin{picture}(180,60)
\put(-45,10){A}
\thicklines \multiput(10,10)(40,0){3}{\circle{10}}
\multiput(15,10)(40,0){2}{\line(1,0){30}}
\put(50,50){\circle{10}} \put(50,15){\line(0,1){30}}
\put(50,-30){\circle{10}} \put(50,5){\line(0,-1){30}}
\put(5,-5){$\alpha_1$} \put(30,-5){$\alpha_2$}
 \put(30,-30){$\alpha_4$}  \put(30,45){$\alpha_5$}\put(85,-5){$\alpha_3$}
\end{picture}
}
\end{center}

\begin{center}
\scalebox{.5}{
\begin{picture}(180,60)
\put(-45,10){B}
\thicklines \multiput(10,10)(40,0){2}{\circle{10}}
\put(15,10){\line(1,0){30}}
\put(70,30){\circle{10}} \put(55,15){\line(1,1){10}}
\put(70,-10){\circle{10}} \put(55,5){\line(1,-1){10}}
\put(90,10){\circle{10}} \put(85,15){\line(-1,1){10}}
\put(85,5){\line(-1,-1){10}}
\put(5,-5){$\alpha_1$} \put(30,-5){$\alpha_2$}
 \put(80,-15){$\alpha_4$}  \put(80,30){$\alpha_5$}\put(95,0){$\alpha_3$}
\end{picture}
}
\end{center}

\begin{center}
\scalebox{.5}{
\begin{picture}(180,60)
\put(-45,10){C}
\thicklines \multiput(10,10)(40,0){4}{\circle{10}}
\multiput(15,10)(40,0){3}{\line(1,0){30}}
\put(50,50){\circle{10}} \put(50,15){\line(0,1){30}}
\put(5,-5){$\alpha_1$} \put(45,-5){$\alpha_2$}
 \put(125,-5){$\alpha_5$}  \put(30,45){$\alpha_3$}\put(85,-5){$\alpha_4$}
\end{picture}
} \end{center}

\begin{center}
\scalebox{.5}{
\begin{picture}(180,60)
\put(-45,10){D}
\thicklines \multiput(10,10)(40,0){4}{\circle{10}}
\multiput(15,10)(40,0){3}{\line(1,0){30}}
\put(50,50){\circle{10}} \put(50,15){\line(0,1){30}}
\put(5,-5){$\alpha_4$} \put(45,-5){$\alpha_3$}
 \put(125,-5){$\alpha_1$}  \put(30,45){$\alpha_5$}\put(85,-5){$\alpha_2$}
\end{picture}
} \end{center}
Note that C and D simply differ by the assignment of the symmetry
root.
\newline

\begin{center}
\scalebox{.5}{
\begin{picture}(180,60)
\put(-45,10){E}
\thicklines \multiput(10,10)(40,0){5}{\circle{10}}
\multiput(15,10)(40,0){4}{\line(1,0){30}}
\put(5,-5){$\alpha_1$} \put(45,-5){$\alpha_2$}
 \put(125,-5){$\alpha_4$}  \put(165,-5){$\alpha_5$}\put(85,-5){$\alpha_3$}
\end{picture}
} \end{center}

\textbf{Case A} - This case may be discarded. Indeed, there are in
fact two hyperbolic algebras with a Dynkin diagram of that shape:
one of them has a long and four short roots, while the other one
has one short and four long roots. Either one cannot find
couplings that reproduce their Cartan matrix or it is the
subdominant condition that is violated. More concretely:
\newline

A.1. Consider first the case for which
$\alpha_1,\alpha_2,\alpha_3,\alpha_4$ correspond to the short roots and
$\alpha_5$ is the long root. Then, according to \cite{S}, one needs
\be
\Vert\alpha_1\Vert^2=\Vert\alpha_2\Vert^2 =\Vert\alpha_3\Vert^2 =
\Vert\alpha_4\Vert^2= 2 \quad\mbox{and}\quad \Vert\alpha_5\Vert^2=4.\ee
These conditions are immediately translated into \be
\lambda^2=2\quad,\quad
\lambda^{\prime 2}+\mu^{\prime 2}=2\quad,\quad \lambda^{\prime\prime
2}+\mu^{\prime\prime 2} + \nu^{\prime\prime 2}=2\quad,\quad
\lambda^{\prime\prime\prime 2}+\mu^{\prime\prime\prime 2} +
\nu^{\prime\prime\prime 2}=4;\ee Hence $\lambda=\sqrt{2}$. From the shape of
the diagram or equivalently from the elements of the Cartan matrix, one
infers successively
\begin{enumerate}
\item{$A_{23}=-1=-\lambda\lambda'$} which gives $\lambda' = 1/\sqrt{2}$ and
$\mu'=\sqrt{3/2}$;
\item{$A_{24}=-1=-\lambda\lambda''$ and
$A_{34}=0=\lambda'\lambda''-\mu'\mu''$} which gives
$\lambda''= 1/\sqrt{2}$, $\mu''=1/\sqrt{6}$ and $\nu''= 2/\sqrt{3}$
\item{$A_{25}=-2=-\lambda\lambda'''$} which gives $\lambda'''=\sqrt{2}$
\item{$A_{35}=0=\lambda'\lambda'''-\mu'\mu'''$} which gives
$\mu'''=\sqrt{2/3}$ and, using the norm of $\alpha_5$,
$\nu'''=2/\sqrt{3}$.
\end{enumerate}
Notice that the condition $A_{45}=0=\lambda''\lambda'''+\mu''\mu'''
-\nu''\nu'''$ is identically satisfied.

In summary, in order to fit the Dynkin diagram displayed in A
(with simple lines between $\alpha_2$ and
$\alpha_1,\alpha_3,\alpha_4$ and a double line between $\alpha_2$
and $\alpha_5$ oriented towards $\alpha_2$), besides the symmetry
wall, we need the following set of dominant walls
\begin{eqnarray}
\alpha_2 = \beta^1-\sqrt{2}\phi\quad\quad &,&\quad \alpha_4 =
\frac{\phi}{\sqrt{2}} + \frac{\varphi}{\sqrt{6}}
-\frac{2\,\psi}{\sqrt{3}} \\ \alpha_3 =
\frac{\phi}{\sqrt{2}}-\sqrt{\frac{3}{2}}\,\varphi\quad &,&\quad
\alpha_5 = \sqrt{2}\phi + \sqrt{\frac{2}{3}}\,\varphi
+\frac{2\,\psi}{\sqrt{3}} \end{eqnarray} It is now easy to verify,
for instance, that $\tilde\alpha_3 =
\beta^1-\frac{\phi}{\sqrt{2}}+\sqrt{\frac{3}{2}}\,\varphi$ cannot
be written as a positive linear combination of the $\alpha_i,
i=2,...,5$. Accordingly, on account of the subdominant conditions,
this case has to be rejected.
\newline

A.2.  There is another possibility producing the same diagram as
in A.1. above where the symmetry wall $\alpha_1$ now plays the
r\^ole of the long root: their norms are \be
\Vert\alpha_1\Vert^2=2\quad\mbox{and}\quad
\Vert\alpha_2\Vert^2=\Vert\alpha_3\Vert^2 =\Vert\alpha_4\Vert^2 =
\Vert\alpha_5\Vert^2= 1 \ee but the equations giving the couplings
analogous to eq.1. to eq.4. above have no solution.
\newline

A.3. In the third case, there a short and four long roots with
norms \be \Vert\alpha_1\Vert^2=\Vert\alpha_2\Vert^2
=\Vert\alpha_3\Vert^2 = \Vert\alpha_4\Vert^2= 2
\quad\mbox{and}\quad \Vert\alpha_5\Vert^2=1.\ee One can solve the
equations for the couplings and write the following set of
billiard walls: the symmetry wall $\alpha_1=\beta^2-\beta^1$ and
\begin{eqnarray}
\alpha_2 =
\beta^1-\sqrt{2}\phi\quad\quad &,&\quad \alpha_4 = \frac{\phi}{\sqrt{2}} +
\frac{\varphi}{\sqrt{6}} -\frac{2\,\psi}{\sqrt{3}} \\ \alpha_3 =
\frac{\phi}{\sqrt{2}}-\sqrt{\frac{3}{2}}\,\varphi\quad &,&\quad \alpha_5 =
\frac{\phi}{\sqrt{2}} +
\frac{\varphi}{\sqrt{6}} +\frac{\psi}{\sqrt{3}}. \end{eqnarray}
However, like in case A.1. above, one sees immediately that
$\tilde\alpha_3 = \beta^1 -
\frac{\phi}{\sqrt{2}} +
\frac{3\varphi}{\sqrt{6}}$, for instance, is not
subdominant; that is the reason why we discard this possibility.
\newline

\textbf{Cases B} - There are three hyperbolic algebras with a Dynkin
diagram of this shape.

B.1. The first one admits the following
couplings

\begin{eqnarray}
\lambda &=& \sqrt{2} \ ; \ \lambda' ={1\over \sqrt{2}} \ ; \ \mu' = \sqrt{{3
\over 2}} \nonumber
\nonumber\\
\lambda''&=& 0\quad \ ; \ \mu'' = \sqrt{{2 \over 3}}  \ ; \ \nu'' = {2
\over
\sqrt{3}}\ ; \
\lambda''' ={1\over \sqrt{2}} \ ; \ \mu''' = {1 \over \sqrt{6}}
\ ; \ \nu''' = {2 \over \sqrt{3} }
\end{eqnarray}

and is the overextension
$A_3^{\wedge\wedge}$

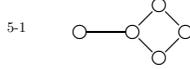
\begin{center}
\scalebox{.5}{
\begin{picture}(180,60)
\put(-45,10){5-1}
\thicklines \multiput(10,10)(40,0){2}{\circle{10}}
\put(15,10){\line(1,0){30}}
\put(70,30){\circle{10}} \put(55,15){\line(1,1){10}}
\put(70,-10){\circle{10}} \put(55,5){\line(1,-1){10}}
\put(90,10){\circle{10}} \put(85,15){\line(-1,1){10}}
\put(85,5){\line(-1,-1){10}}
\end{picture}
}
\end{center}

B.2. The second one has the
following Dynkin diagram

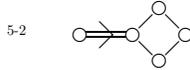
\begin{center}
\scalebox{.5}{
\begin{picture}(180,60)
\put(-45,10){5-2}
 \thicklines
 \put(15,8){\line(1,0){30}} \put(15,12){\line(1,0){30}}
\multiput(10,10)(40,0){2}{\circle{10}} \put(25,0){\line(1,1){10}}
\put(25,20){\line(1,-1){10}}
\put(70,30){\circle{10}} \put(55,15){\line(1,1){10}}
\put(70,-10){\circle{10}} \put(55,5){\line(1,-1){10}}
\put(90,10){\circle{10}} \put(85,15){\line(-1,1){10}}
\put(85,5){\line(-1,-1){10}}
\end{picture}
}
\end{center}

and the following set of dilaton couplings:

\begin{eqnarray}
\lambda &=& 1\ ; \
\lambda' = {1\over 2} \ ; \ \mu' = {\sqrt{3} \over 2} \nonumber \\
\lambda''&=& 0 \ ; \ \mu'' = {1 \over \sqrt{3}}  \ ; \ \nu'' =
{\sqrt{2}
 \over \sqrt{3}} \ ; \
\lambda''' = {1\over 2} \ ; \ \mu''' = {1 \over 2\sqrt{3}} \ ; \
\nu''' = { \sqrt{2} \over \sqrt{3} }.
\end{eqnarray}

B.3. The diagram of the third one is the same as $(5-2)$ but with
the reversed arrow: this is impossible since in the present
context the norms are required to satisfy
$\Vert\alpha_1\Vert^2\geq \Vert\alpha_2\Vert^2$.
\newline

\textbf{Case C} - In order to generate this kind of structure, one
needs $\lambda'''=\mu''' =0$ and $\lambda' \lambda'' = \mu'
\mu''$. Next, from the subdominant condition for $\tilde\alpha_3$,
we deduce that $A_{32}$ can be $-2$ or $-3$ but since we want
hyperbolic algebras, only the value $A_{32} = -2$ can be retained.
Therefore $A_{23} = -1$ and $\lambda=\sqrt{2}$, $\lambda'=
\lambda''= 1 / \sqrt{2}$, $\mu'=1 / \sqrt{2}$, $\mu''= 1 /
\sqrt{2}$ and $\nu''= 1$.  A priori, one might still have
$\nu'''=2,1,\sqrt{2}$ but only one value is compatible with the
magnetic wall $\tilde\alpha_5$ being subdominant, namely $\nu''' =
1$. Accordingly, the couplings need to be defined as

\begin{eqnarray}
\lambda &=& \sqrt{2} \ ; \
\lambda' = {1\over \sqrt{2}} \ ; \ \mu' = {1\over \sqrt{2}} \nonumber
\\
\lambda''&=& {1\over \sqrt{2}} \ ; \ \mu'' = {1\over \sqrt{2}}  \ ; \
\nu'' =1 \ ; \
\lambda''' = 0\quad \ ; \ \mu''' = 0\quad \ ; \ \nu''' = 1 .
\end{eqnarray}

and the Dynkin diagram is the following,

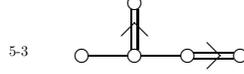
\begin{center}
\scalebox{.5}{
\begin{picture}(180,60)
\put(-45,10){5-3}
\thicklines \multiput(10,10)(40,0){4}{\circle{10}}
\multiput(15,10)(40,0){2}{\line(1,0){30}}
\put(95,7.5){\line(1,0){30}} \put(95,12.5){\line(1,0){30}}
\put(105,0){\line(1,1){10}} \put(105,20){\line(1,-1){10}}
\put(50,50){\circle{10}} \put(52.5,15){\line(0,1){30}}
\put(47.7,15){\line(0,1){30}}
\put(40,25){\line(1,1){10}} \put(50,35){\line(1,-1){10}}
\end{picture}
} \end{center}

\textbf{Cases D} - Dynkin diagrams of this shape can only be
recovered with
\be \lambda'' =
\lambda''' = 0\quad\mbox{and either}\quad \lambda =
\sqrt{2}\quad\mbox{or}\quad
\lambda = 1.\ee

\textbf{D.1.
$\lambda =
\sqrt{2}$}
\newline

All hyperbolic diagrams of that type have in their Cartan matrix
$A_{34} = A_{43} = -1$ which means that
$\Vert\alpha_3\Vert^2=\Vert\alpha_4\Vert^2$. Two additional cases
must be considered depending on which of $\alpha_2$ or $\alpha_5$
has a norm equal to the norm of $\alpha_3$:
\begin{enumerate}
\item in case D.1.1. we assume that the norms of
$\alpha_3$,
$\alpha_4$ and
$\alpha_5$ are equal
\item in case D.1.2. we assume that the norms of $\alpha_2$,
$\alpha_3$ and
$\alpha_4$ are equal.
\end{enumerate}
The subdominant conditions here simply reduce to
$A_{23}=-1$.
\newline

D.1.1. Again two hyperbolic algebras correspond to this case. For the
first one, the billiard walls are built out of the following
couplings

\begin{eqnarray}
\lambda &=& \sqrt{2}  \ ; \
\lambda'= {1\over \sqrt{2}} \ ; \ \mu' = \sqrt{{3\over 2}} \nonumber
\\
\lambda''&=& 0 \quad \ ; \ \mu'' = \sqrt{{2\over 3}}  \ ; \ \nu''
={1\over\sqrt{3}}  \ ; \
\lambda''' = 0\quad \ ; \ \mu''' = \sqrt{{2\over 3}} \ ; \
\nu''' = {2 \over \sqrt{3 }}\label{435}
\end{eqnarray}

and the Dynkin diagram is that of the overextension $B_3^{\wedge\wedge}$

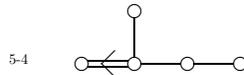
\begin{center}
\scalebox{.5}{
\begin{picture}(180,60)
\put(-45,10){5-4}
\thicklines \multiput(10,10)(40,0){4}{\circle{10}}
\multiput(55,10)(40,0){2}{\line(1,0){30}}
\put(15,8){\line(1,0){30}} \put(15,12){\line(1,0){30}}
\put(25,10){\line(1,1){10}} \put(25,10){\line(1,-1){10}}
\put(50,50){\circle{10}} \put(50,15){\line(0,1){30}}
\end{picture}
} \end{center}

One can produce a billiard for the second one using the same
couplings as in (\ref{435}) except for

\begin{eqnarray}
\lambda''&=& 0 \ ; \ \mu'' = 2 \sqrt{{2\over 3}}  \ ; \ \nu''
={2\over\sqrt{3}}. \end{eqnarray}

The Dynkin diagram here describes the twisted overextension
$A_5^{(2)\wedge}$

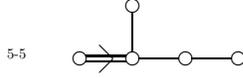
\begin{center}
\scalebox{.5}{
\begin{picture}(180,60)
\put(-45,10){5-5} \thicklines
\multiput(10,10)(40,0){4}{\circle{10}}
\multiput(55,10)(40,0){2}{\line(1,0){30}}
\put(15,8){\line(1,0){30}} \put(15,12){\line(1,0){30}}
\put(25,0){\line(1,1){10}} \put(25,20){\line(1,-1){10}}
\put(50,50){\circle{10}} \put(50,15){\line(0,1){30}}
\end{picture}
} \end{center}

D.1.2.  Here also, two hyperbolic algebras correspond to this case
but one is eliminated on account of the subdominant conditions.
For the remaining one, the couplings are

\begin{eqnarray}
\lambda &=& \sqrt{2}  \ ; \
\lambda' ={1\over \sqrt{2}} \ ; \ \mu' = {1\over \sqrt{2}} \nonumber
\\
\lambda''&=& 0 \ ; \ \mu'' = {1\over \sqrt{2}}  \ ; \ \nu'' ={1\over
\sqrt{2}} \ ; \
\lambda''' = 0 \ ; \ \mu''' ={1\over \sqrt{2}} \ ; \ \nu'''
= {1\over \sqrt{2}}
\end{eqnarray}
and the Dynkin diagram is

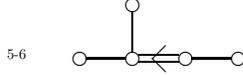
\begin{center}
\scalebox{.5}{
\begin{picture}(180,60)
\put(-45,10){5-6 }
\thicklines \multiput(10,10)(40,0){4}{\circle{10}}
\put(15,10){\line(1,0){30}} \put(95,10){\line(1,0){30}}
\put(55,12.5){\line(1,0){30}} \put(55,7.5){\line(1,0){30}}
\put(65,10){\line(1,1){10}} \put(65,10){\line(1,-1){10}}
\put(50,50){\circle{10}} \put(50,15){\line(0,1){30}}
\end{picture}
} \end{center}

\textbf{D.2. $\lambda = 1$}
\newline

In table 2 of reference \cite{S}, there are
two hyperbolic algebras with a Dynkin diagram of this shape. Both are
admissible for our present purpose:

D.2.1. The first one has couplings given by

\begin{eqnarray}
\lambda &=& 1  \ ; \
\lambda' = {1\over 2} \ ; \ \mu' = { \sqrt{3}\over 2} \nonumber \\
\lambda'' &=& 0 \ ; \ \mu'' ={2\over \sqrt{3}} \ ; \ \nu'' =
\sqrt{{2\over 3}} \ ; \
\lambda'''= 0 \ ; \ \mu''' = {1\over \sqrt{3}}  \ ; \
\nu''' =  \sqrt{{2\over3}}
\end{eqnarray}

and corresponds to
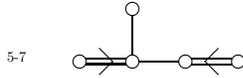
\begin{center}
\scalebox{.5}{
\begin{picture}(180,60)
\put(-45,10){5-7} \thicklines
\multiput(10,10)(40,0){4}{\circle{10}} \put(55,10){\line(1,0){30}}
\put(15,8){\line(1,0){30}} \put(15,12){\line(1,0){30}}
\put(25,0){\line(1,1){10}} \put(25,20){\line(1,-1){10}}
\put(50,50){\circle{10}} \put(50,15){\line(0,1){30}}
\put(95,12.5){\line(1,0){30}} \put(95,7.5){\line(1,0){30}}
\put(105,10){\line(1,1){10}} \put(105,10){\line(1,-1){10}}
\end{picture}
} \end{center}
D.2.2. The second one requires

\begin{eqnarray}
\lambda &=& 1  \ ; \
\lambda' = {1\over 2} \ ; \ \mu' = { \sqrt{3}\over 2} \nonumber \\
\lambda'' &=& 0 \ ; \ \mu'' ={1\over \sqrt{3}} \ ; \ \nu'' =
{1\over \sqrt{6}} \ ; \
\lambda'''= 0 \ ; \ \mu''' = {1\over \sqrt{3}}  \ ; \
\nu''' =  \sqrt{{2\over3}}
\end{eqnarray}
and has the following diagram

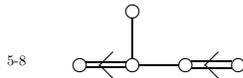
\begin{center}
\scalebox{.5}{
\begin{picture}(180,60)
\put(-45,10){5-8} \thicklines
\multiput(10,10)(40,0){4}{\circle{10}} \put(55,10){\line(1,0){30}}
\put(15,8){\line(1,0){30}} \put(15,12){\line(1,0){30}}
\put(25,10){\line(1,1){10}} \put(25,10){\line(1,-1){10}}
\put(50,50){\circle{10}} \put(50,15){\line(0,1){30}}
\put(95,12.5){\line(1,0){30}} \put(95,7.5){\line(1,0){30}}
\put(105,10){\line(1,1){10}} \put(105,10){\line(1,-1){10}}
\end{picture}
} \end{center}

\textbf{Cases E} - Table 2 of reference \cite{S} displays two
hyperbolic algebras of rank 5 which are duals of each other and
have linear diagrams. Only one of these two can be associated to a
billiard the walls of which correspond to

E.1. $\lambda =
\sqrt{2}$, $\lambda'' =
\lambda'''=
\mu''' =0$  and all other dilaton couplings equal
to
$1/\sqrt{2}$.

Its Dynkin diagram is the twisted overextension
$A_{6}^{(2)\wedge}$ and is given by

\begin{center}
\scalebox{.5}{
\begin{picture}(180,60)
\put(-45,10){5-9}
\thicklines \multiput(10,10)(40,0){5}{\circle{10}}
\put(15,10){\line(1,0){30}} \put(95,10){\line(1,0){30}}
\put(55,8){\line(1,0){30}} \put(55,12){\line(1,0){30}}
\put(65,0){\line(1,1){10}} \put(65,20){\line(1,-1){10}}
\put(135,8){\line(1,0){30}} \put(135,12){\line(1,0){30}}
\put(145,0){\line(1,1){10}} \put(145,20){\line(1,-1){10}}
\end{picture}
} \end{center}

There are however two more hyperbolic algebras with such linear Dynkin
diagrams; they are missing in \cite{S} but perfectly relevant in
the present context:

E.2. the first one is the overextension
$C_3^{\wedge\wedge}$
\begin{center}
\scalebox{.5}{
\begin{picture}(180,60)
\put(-45,10){5-10}
\thicklines \multiput(10,10)(40,0){5}{\circle{10}}
\put(15,10){\line(1,0){30}} \put(95,10){\line(1,0){30}}
\put(55,8){\line(1,0){30}} \put(55,12){\line(1,0){30}}
\put(65,0){\line(1,1){10}} \put(65,20){\line(1,-1){10}}
\put(135,8){\line(1,0){30}} \put(135,12){\line(1,0){30}}
\put(145,10){\line(1,1){10}} \put(145,10){\line(1,-1){10}}
\end{picture}
} \end{center}
whose couplings are equal to the previous ones except $\nu'''=\sqrt{2}$.

E.3. The second one is the dual of $C_3^{\wedge\wedge}$ known as
the twisted overextension $D_4^{(2)\wedge}$; its Dynkin diagram
corresponds to the previous one with reversed arrows

\begin{center}
\scalebox{.5}{
\begin{picture}(180,60)
\put(-45,10){5-11}
\thicklines \multiput(10,10)(40,0){5}{\circle{10}}
\put(15,10){\line(1,0){30}} \put(95,10){\line(1,0){30}}
\put(55,8){\line(1,0){30}} \put(55,12){\line(1,0){30}}
\put(65,10){\line(1,1){10}} \put(65,10){\line(1,-1){10}}
\put(135,8){\line(1,0){30}} \put(135,12){\line(1,0){30}}
\put(145,0){\line(1,1){10}} \put(145,20){\line(1,-1){10}}
\end{picture}
} \end{center}
and the dilaton couplings are such that $\lambda = \lambda'=\mu' = \mu''
= \nu'' = \nu''' = \sqrt{2}$ while $\lambda'' = \lambda''' = \mu'''=0$.

\subsubsection{Two or more magnetic walls}

That these cases may be discarded will be proved on a particular
case but the argument can be easily generalized. Suppose the
dominant set comprises two magnetic walls and two electric ones,
we can always choose the parametrization such that
\begin{eqnarray} \alpha_1 &=& \beta^2-\beta^1\nonumber\\ \alpha_2 &=&
\beta^1 -\lambda\,\phi -\mu\,\varphi -\nu\,\psi\nonumber\\ \alpha_3 &=&
\beta^1 -\lambda'\,\phi -\mu'\,\varphi +\nu'\,\psi \label{550}\\ \alpha_4
&=&
\lambda''\,\phi +\mu''\,\varphi \nonumber\\ \alpha_5 &=&
\lambda'''\,\phi.\nonumber
\end{eqnarray}
Being assumed subdominant, the electric walls $\tilde\alpha_2 =
\lambda\,\phi +\mu\,\varphi +\nu\,\psi$ and $\tilde\alpha_3 =
\lambda'\,\phi +\mu'\,\varphi -\nu'\,\psi$, independent of
$\beta^1$, must be written as positive linear combinations of
$\alpha_4$ and $\alpha_5$ only; this requires $\nu=\nu'=0$ but
then (\ref{550}) can no longer describe a rank five root system.
\newline

The same argument remains of course valid for more magnetic walls.

\subsection{$D>3$}

The empirical oxidation rule set up in the previous sections also
holds for the rank 5 algebras:
\newline

\begin{enumerate}
\item{Diagram $(5-1)$} is the Dynkin diagram of the overextension
$A_3^{\wedge\wedge}$. The Lagrangian is that of pure gravity in
$D_{max}=6$.
\item{Diagram $(5-2)$} : the 3-dimensional Lagrangian
cannot be oxidized because of the norm of $\alpha_2$.
\item{Diagram $(5-3)$} : The Lagrangian can be oxidized twice,
up to $D_{max}=5$. The dominant walls are the three symmetry walls
$\alpha_1 =
\beta^4-\beta^3$,
$\alpha_2=\beta^3-\beta^2$, $\alpha_3=\beta^2-\beta^1$ and the
electric wall of a $1$-form \be \alpha_4 = \beta^1 - \sqrt{1/3}  \phi
\ee and its magnetic wall\be \alpha_5 =
\beta^1 +
\beta^2 +
\sqrt{1/3}  \phi.\ee Obviously, $\tilde\alpha_4=\alpha_5$ and
$\tilde\alpha_5=\alpha_4$.
\item{Diagram $(5-4)$} represents $B_3^{\wedge\wedge}$. The
maximally oxidized Lagrangian is six-dimensional. The dominant walls
are here the four symmetry walls
$\alpha_1 =
\beta^5-\beta^4$, $\alpha_2=\beta^4-\beta^3$,
$\alpha_3=\beta^3-\beta^2$, $\alpha_4 = \beta^2-\beta^1$ and
\be\alpha_5 = \beta^1 + \beta^2\ee which is the electric
or magnetic wall of a self-dual $2-$form:
obviously $\tilde\alpha_5=\alpha_5$.
\item{Diagram $(5-5)$} is the twisted overextension
$A_5^{(2)\wedge}$. Here, $D_{max}=4$ and besides the symmetry
walls $\alpha_1=\beta^3-\beta^2$ and $\alpha_2= \beta^2-\beta^1$,
one finds
\begin{eqnarray} \alpha_3 &=& \beta^1-\sqrt{3/2} \phi\\ \alpha_4 &=&
\sqrt{2/3}\phi - 2/
\sqrt{3} \varphi\\ \alpha_5 &=& 2 \sqrt{2/3} \varphi + 2/ \sqrt{3}
\psi\end{eqnarray} which are the electric walls respectively of a one-form
and two
$0$-forms. One easily checks that $\tilde\alpha_3 =
\alpha_3+\alpha_4+\alpha_5$, $\tilde\alpha_4=
\alpha_2+2\alpha_3+\alpha_5$ and
$\tilde\alpha_5=\alpha_2+2\alpha_3+\alpha_4.$
\item{Diagram $(5-6)$} : $D_{max}=4$ and one needs
$\alpha_1=\beta^3-\beta^2$, $\alpha_2= \beta^2-\beta^1$ and
\begin{eqnarray}\alpha_3 &=& \beta^1-\sqrt{1/2} \phi\\ \alpha_4 &=&
\sqrt{1/2}\phi - \sqrt{1/2} \varphi\\\alpha_5 &=& \sqrt{1/2} \phi
+ \sqrt{1/2} \varphi.\end{eqnarray} The form-field content is the
same as the previous one but the dilaton couplings are different.
Moreover: $\tilde\alpha_3=\alpha_3+\alpha_4+\alpha_5$,
$\tilde\alpha_4= \alpha_2+2\alpha_3+\alpha_5$ and $\tilde\alpha_5=
\alpha_2+2\alpha_3+\alpha_4$.
\item{Diagrams $(5-7)$ and $(5-8)$} : their 3-D Lagrangians cannot be
further oxidized because of the norm of $\alpha_2$.
\item{Diagram $(5-9)$} describes $A_{6}^{(2)\wedge}$. Here, $D_{max}=4$.
One obtains the billiard with $\alpha_1=\beta^3-\beta^2$,
$\alpha_2= \beta^2-\beta^1$ and\begin{eqnarray}\alpha_3 &=&
\beta^1-\sqrt{1/2} \phi\\ \alpha_4 &=& \sqrt{1/2}\phi - \sqrt{1/2}
\varphi \\\alpha_5 &=& \sqrt{1/2} \varphi.\end{eqnarray} One draws
the same conclusion as for $(5-6)$ and $(5-9)$ above. Here again:
$\tilde\alpha_3=\alpha_3+\alpha_4+\alpha_5$, $\tilde\alpha_4=
\alpha_2+2\alpha_3+\alpha_4+2\alpha_5$ and $\tilde\alpha_5=
\alpha_2+2\alpha_3+2\alpha_4+\alpha_5$.
\item{Diagram $(5-10)$} is the overextension
$C_3^{\wedge\wedge}$; the maximal oxidation dimension is $D_{max}=
4$ and the corresponding Lagrangian can be found in \cite{DdBHS}.
\item{Diagram $(5-11)$} is the twisted overextension $D_4^{(2)\wedge}$. No
Lagrangian exists in higher dimensions.
\end{enumerate}

\textbf{Comment}\newline

The results of this section show again that the subdominant
conditions play an important r\^ole in three dimensions where they
effectively contribute to the elimination of several Dynkin
diagrams. However, once they are satisfied in three dimensions,
they are always fulfilled in all dimensions where a Lagrangian
exists and only integers enter the linear combinations.

\section{Rank 6 Hyperbolic algebras}

\subsection{$D=3$}

The number of dilatons in the 3-dimensional Lagrangian is equal to
$N=4$: we denote them by $\phi^1=\phi, \phi^2=\varphi,
\phi^3=\psi, \phi^4=\chi$. A straightforward generalization of the
argument used in the previous sections implies that a single
configuration for the set of dominant walls has to be considered.
It comprises one magnetic wall and four electric ones.\newline

After allowed simplifications, the dominant walls are parametrized
according to:
\begin{eqnarray}\alpha_1 &=&\beta^2-\beta^1,\\
\alpha_2 &=& \beta^1 - \lambda \phi, \\ \alpha_3 &=&\lambda'\phi - \mu'
\varphi,\\ \alpha_4 &=&\lambda'' \phi + \mu''\varphi - \nu''
\psi,\\ \alpha_5 &=& \lambda''' \phi + \mu''' \varphi + \nu'''
\psi - \rho''' \chi,\\
\alpha_6 &=& \lambda'''' \phi +
\mu'''' \varphi + \nu'''' \psi + \rho''''\chi.\end{eqnarray}

The structure of the Dynkin diagrams is therefore displayed in one
of the cases labelled $A$ to $E$ below, depending on the number of
vertices connected to $\alpha_2$. When necessary, further
subclasses are introduced according to the number of vertices
linked to $\alpha_3$,...
\newline

\textbf{Case A} - The central vertex is labelled $\alpha_2$ and is
connected to the five other vertices:

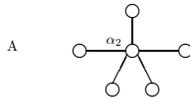
\begin{center}
\scalebox{.5}{
\begin{picture}(180,60)
\put(-45,10){A}
\thicklines \multiput(10,10)(40,0){3}{\circle{10}}
\multiput(15,10)(40,0){2}{\line(1,0){30}}
\put(50,40){\circle{10}} \put(50,15){\line(0,2){20}}
\put(30,15){$\alpha_2$}
\put(35,-20){\circle{10}} \put(65,-20){\circle{10}}
\put(65,-15){\line(-1,2){11}} \put(35,-15){\line(1,2){11}}
\end{picture}
} \end{center}

There is a single hyperbolic algebra of this type in \cite{S}; one can
solve the equations for the dilaton couplings, but the
subdominant walls are not expressible as positive linear combinations of
the dominant ones.\newline

\textbf{Case B} - The root $\alpha_2$ is connected to four vertices:

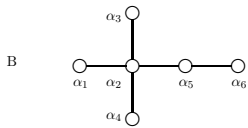
\begin{center}
\scalebox{.5}{
\begin{picture}(180,60)
\put(-45,10){B}
\thicklines \multiput(10,10)(40,0){4}{\circle{10}}
\multiput(15,10)(40,0){3}{\line(1,0){30}}
\put(50,-30){\circle{10}} \put(50,5){\line(0,-1){30}}
\put(50,50){\circle{10}} \put(50,15){\line(0,1){30}}
\put(5,-5){$\alpha_1$} \put(30,-5){$\alpha_2$}
 \put(125,-5){$\alpha_6$}  \put(30,45){$\alpha_3$}\put(85,-5){$\alpha_5$}
 \put(30,-30){$\alpha_4$}
\end{picture}
} \end{center}

There are three hyperbolic algebras with that kind of Dynkin
diagram but none of them can be retained: indeed, couplings exist
but the subdominant conditions cannot be fulfilled.
\newline

\textbf{Cases C} - $\alpha_2$ has three links.\newline

One has first the loop diagram

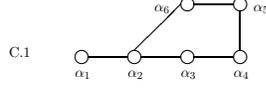
\begin{center}
\scalebox{.5}{
\begin{picture}(180,60)
\put(-45,10){C.1}
\thicklines \multiput(10,10)(40,0){4}{\circle{10}}
\multiput(15,10)(40,0){3}{\line(1,0){30}}
\multiput(90,50)(40,0){2}{\circle{10}}
\put(130,15){\line(0,1){30}} \put(50,15){\line(1,1){35}}
\put(95,50){\line(1,0){30}}
\put(5,-5){$\alpha_1$} \put(45,-5){$\alpha_2$}
 \put(125,-5){$\alpha_4$}  \put(65,45){$\alpha_6$}\put(85,-5){$\alpha_3$}
  \put(140,45){$\alpha_5$}
\end{picture}
} \end{center}

One hyperbolic algebra has such a Dynkin diagram, namely, the
overextension $A_4^{\wedge\wedge}$. As we already know from
\cite{DHN}, the searched for $3$-dimensional Lagrangian coincides
with the toroidal dimensional reduction of the seven-dimensional
Einstein-Hilbert Lagrangian. The dilaton couplings are given by

\begin{eqnarray}
\lambda &=& \sqrt{2} \ ; \
\lambda' = {1\over \sqrt{2}} \ ; \ \mu' =  \sqrt{{3\over 2}} \ ; \
\lambda''= 0 \ ; \ \mu'' = \sqrt{{2\over 3}}  \ ; \ \nu'' =  { 2
\over
\sqrt{3}} \ ; \
\lambda''' = 0 \ ; \ \mu''' =0;  \nonumber \\ \nu''' &=& { \sqrt{3} \over
2} \ ; \ \rho'''={ \sqrt{5} \over 2} \ ; \
\lambda''''={1\over \sqrt{2}}  \ ; \ \mu''''={1\over
\sqrt{6}} \ ; \ \nu''''={1\over 2 \sqrt{3}}  \ ; \
\rho''''={ \sqrt{5} \over 2}
\end{eqnarray}

and its Dynkin diagram is

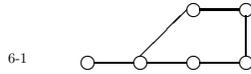
\begin{center} \scalebox{.5}{
\begin{picture}(180,60)
\put(-50,10){6-1}
\thicklines \multiput(10,10)(40,0){4}{\circle{10}}
\multiput(15,10)(40,0){3}{\line(1,0){30}}
\multiput(90,50)(40,0){2}{\circle{10}}
\put(130,15){\line(0,1){30}} \put(50,15){\line(1,1){35}}
\put(95,50){\line(1,0){30}}
\end{picture}
} \end{center}

Next comes the tree diagram

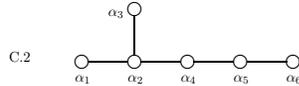
\begin{center}
\scalebox{.5}{
\begin{picture}(180,60)
\put(-45,10){C.2}
\thicklines \multiput(10,10)(40,0){5}{\circle{10}}
\multiput(15,10)(40,0){4}{\line(1,0){30}}
\put(50,50){\circle{10}} \put(50,15){\line(0,1){30}}
\put(5,-5){$\alpha_1$} \put(45,-5){$\alpha_2$}
 \put(125,-5){$\alpha_5$}  \put(30,45){$\alpha_3$}\put(85,-5){$\alpha_4$}
  \put(165,-5){$\alpha_6$}
\end{picture}
} \end{center}

Two hyperbolic algebras have a Dynkin diagram of this shape;  but
they cannot be associated to billiards again because of the
impossibility to satisfy the subdominant conditions.
\newline

One also has to allow a relabelling of the vertices according to

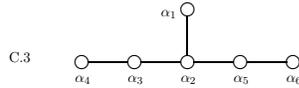
\begin{center}
\scalebox{.5}{
\begin{picture}(180,60)
\put(-45,10){C.3}
\thicklines \multiput(10,10)(40,0){5}{\circle{10}}
\multiput(15,10)(40,0){4}{\line(1,0){30}}
\put(90,50){\circle{10}} \put(90,15){\line(0,1){30}}
\put(5,-5){$\alpha_4$} \put(45,-5){$\alpha_3$}
 \put(125,-5){$\alpha_5$}  \put(70,45){$\alpha_1$}\put(85,-5){$\alpha_2$}
  \put(165,-5){$\alpha_6$}
\end{picture}
} \end{center}

There are five hyperbolic algebras of that type; but for only one of
them can one fulfill all conditions. The dilaton couplings are given by

\begin{eqnarray}
\lambda &=& \sqrt{2} \ ; \
\lambda' = {1\over \sqrt{2}} \ ; \ \mu' =  \sqrt{{3\over 2}} \ ; \
\lambda''= 0 \ ; \ \mu'' = \sqrt{{2\over 3}}  \ ; \ \nu'' =  { 1 \over
\sqrt{3}}  \ ; \
\lambda''' = {1\over \sqrt{2}};\nonumber \\ \mu''' &=&{1\over
\sqrt{6}} \ ; \ \nu''' = {1\over \sqrt{3}}  \ ; \ \rho'''=1 \ ; \
\lambda''''= 0 \ ; \ \alpha''''=0 \ ; \ \beta''''=0 \ ; \
\rho''''=1
\end{eqnarray}

and its Dynkin diagram is the following

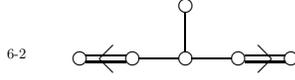
\begin{center}
\scalebox{.5}{
\begin{picture}(180,60)
\put(-45,10){6-2}
\thicklines \multiput(10,10)(40,0){5}{\circle{10}}
\multiput(55,10)(40,0){2}{\line(1,0){30}}
\put(15,7.5){\line(1,0){30}}\put(15,12.5){\line(1,0){30}}
\put(135,7.5){\line(1,0){30}}\put(135,12.5){\line(1,0){30}}
\put(145,0){\line(1,1){10}} \put(145,20){\line(1,-1){10}}
\put(90,50){\circle{10}} \put(90,15){\line(0,1){30}}
\put(25,10){\line(1,1){10}} \put(25,10){\line(1,-1){10}}
\end{picture}
} \end{center}

\textbf{Cases D} are characterized by the fact that $\alpha_2$ has
two links:\newline

D.1. corresponds further to $\alpha_3$ having four links

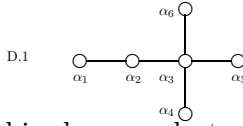
\begin{center}
\scalebox{.5}{
\begin{picture}(180,60)
\put(-45,10){D.1}
\thicklines \multiput(10,10)(40,0){4}{\circle{10}}
\multiput(15,10)(40,0){3}{\line(1,0){30}}
\put(90,50){\circle{10}} \put(90,15){\line(0,1){30}}
\put(90,-30){\circle{10}} \put(90,5){\line(0,-1){30}}
\put(5,-5){$\alpha_1$} \put(45,-5){$\alpha_2$}
 \put(125,-5){$\alpha_5$}  \put(70,45){$\alpha_6$}\put(70,-5){$\alpha_3$}
 \put(70,-30){$\alpha_4$}
\end{picture}
} \end{center}

Three diagrams of \cite{S} fit in this shape; only two of them are
realized through billiards. The couplings of the first one are given by

\begin{eqnarray}
\lambda &=& \sqrt{2} \ ; \
\lambda' = {1\over \sqrt{2}} \ ; \ \mu' =  \sqrt{{3\over 2}} \ ; \
\lambda''= 0 \ ; \ \mu'' = \sqrt{{2\over 3}}= \mu''' \ ; \ \nu'' =  { 2 \over
\sqrt{3}} ;\quad
\lambda''' = 0;\nonumber \\ \nu'''
&=& {1\over
\sqrt{3}};\quad\rho'''=1;\quad
\lambda'''' = 0;\quad\mu''''=\sqrt{{ 2 \over 3}};\quad
\nu''''={1\over \sqrt{3}};\quad
\rho''''=1 \label{560}
\end{eqnarray}

they provide the Dynkin diagram which is $D_4^{\wedge\wedge}$:

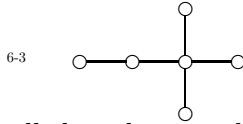
\begin{center}
\scalebox{.5}{
\begin{picture}(180,60)
\put(-45,10){6-3}
\thicklines \multiput(10,10)(40,0){4}{\circle{10}}
\multiput(15,10)(40,0){3}{\line(1,0){30}}
\put(90,50){\circle{10}} \put(90,15){\line(0,1){30}}
\put(90,-30){\circle{10}} \put(90,5){\line(0,-1){30}}
\end{picture}
} \end{center}

For the second one, $\lambda = 1$ and all the other couplings are
those given in (\ref{560}) divided by $\sqrt{2}$. They lead to the
following diagram

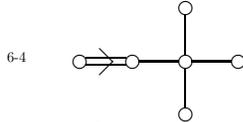
\begin{center}
\scalebox{.5}{
\begin{picture}(180,60)
\put(-45,10){6-4}
\thicklines \multiput(10,10)(40,0){4}{\circle{10}}
\multiput(55,10)(40,0){2}{\line(1,0){30}}
\put(15,7.5){\line(1,0){30}}\put(15,12.5){\line(1,0){30}}
\put(25,0){\line(1,1){10}} \put(25,20){\line(1,-1){10}}
\put(90,50){\circle{10}} \put(90,15){\line(0,1){30}}
\put(90,-30){\circle{10}} \put(90,5){\line(0,-1){30}}
\end{picture}
} \end{center}

Case D.2. corresponds to $\alpha_3$ having three connections

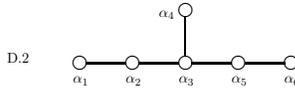
\begin{center}
\scalebox{.5}{
\begin{picture}(180,60)
\put(-45,10){D.2}
\thicklines \multiput(10,10)(40,0){5}{\circle{10}}
\multiput(15,10)(40,0){4}{\line(1,0){30}}
\put(90,50){\circle{10}} \put(90,15){\line(0,1){30}}
\put(5,-5){$\alpha_1$} \put(45,-5){$\alpha_2$}
 \put(125,-5){$\alpha_5$}  \put(70,45){$\alpha_4$}\put(85,-5){$\alpha_3$}
  \put(165,-5){$\alpha_6$}
\end{picture}
} \end{center}

and differs from C.3. above by the assignment of the
symmetry root. There are 4 Dynkin diagrams representing hyperbolic algebras
of this type and they all admit a billiard. \newline

D.2.1. The couplings are

\begin{eqnarray}
\lambda &=& 1 \ ; \
\lambda' = {1\over 2} \ ; \ \mu' =  {\sqrt{3}\over 2} \ ; \
\lambda'' = 0\ ; \ \mu'' = {1\over\sqrt{ 3}}  \ ; \ \nu'' =
\sqrt{{2\over3}} \ ; \
\lambda''' = 0 \ ; \ \mu''' = {1 \over \sqrt{3} } \nonumber \\
\nu''' &=& {1\over \sqrt{6}} \ ; \
\rho'''= {1 \over \sqrt{2}} \ ; \
\lambda'''' = 0 \ ; \ \mu''''=0  \ ; \ \nu''''=0 \ ; \
\rho''''=\sqrt{2}.\label{xx}
\end{eqnarray}

and the Dynkin diagram corresponds to

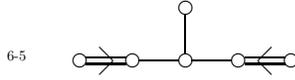
\begin{center}
\scalebox{.5}{
\begin{picture}(180,60)
\put(-45,10){6-5}
\thicklines \multiput(10,10)(40,0){5}{\circle{10}}
\multiput(55,10)(40,0){2}{\line(1,0){30}}
\put(15,7.5){\line(1,0){30}}\put(15,12.5){\line(1,0){30}}
\put(25,0){\line(1,1){10}} \put(25,20){\line(1,-1){10}}
\put(135,7.5){\line(1,0){30}}\put(135,12.5){\line(1,0){30}}
\put(145,10){\line(1,1){10}} \put(145,10){\line(1,-1){10}}
\put(90,50){\circle{10}} \put(90,15){\line(0,1){30}}
\end{picture}
} \end{center}

D.2.2. The couplings are the same as in (\ref{xx}) above except
$\rho''''$ which reads \be \rho'''' = 1/ \sqrt{2}.\ee

The Dynkin diagram is

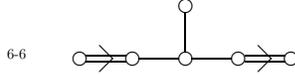
\begin{center}
\scalebox{.5}{
\begin{picture}(180,60)
\put(-45,10){6-6}
\thicklines \multiput(10,10)(40,0){5}{\circle{10}}
\multiput(55,10)(40,0){2}{\line(1,0){30}}
\put(15,7.5){\line(1,0){30}}\put(15,12.5){\line(1,0){30}}
\put(25,0){\line(1,1){10}} \put(25,20){\line(1,-1){10}}
\put(135,7.5){\line(1,0){30}}\put(135,12.5){\line(1,0){30}}
\put(145,0){\line(1,1){10}} \put(145,20){\line(1,-1){10}}
\put(90,50){\circle{10}} \put(90,15){\line(0,1){30}}
\end{picture}
} \end{center}

D.2.3. The dilaton couplings are given by

\begin{eqnarray}
\lambda &=& \sqrt{2} \ ; \
\lambda' = {1\over \sqrt{2}} \ ; \ \mu' =  \sqrt{{3\over 2}} \ ; \
\lambda''= 0= \ ; \ \mu'' = \sqrt{{2\over 3}}  \ ; \ \nu'' =  { 2
\over
\sqrt{3}} \ ; \
\lambda''' = 0;\nonumber \\ \mu''' &=&\sqrt{{ 2 \over 3}}\ ; \ \nu''' =
{1\over
\sqrt{3}}  \ ; \ \rho'''=1 \ ; \
\lambda'''' = 0 \ ; \ \mu''''=0  \ ; \ \nu''''=0 \ ; \
\rho''''=1
 \label{563}
\end{eqnarray}
they provide the Dynkin diagram of $B_4^{\wedge\wedge}$

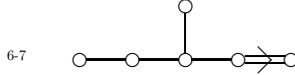
\begin{center}
\scalebox{.5}{
\begin{picture}(180,60)
\put(-45,10){6-7}
\thicklines \multiput(10,10)(40,0){5}{\circle{10}}
\multiput(15,10)(40,0){3}{\line(1,0){30}}
\put(135,7.5){\line(1,0){30}}\put(135,12.5){\line(1,0){30}}
\put(145,0){\line(1,1){10}} \put(145,20){\line(1,-1){10}}
\put(90,50){\circle{10}} \put(90,15){\line(0,1){30}}
\end{picture}
} \end{center}

D.2.4. The couplings are the same as in (\ref{563}) except \be
\rho''''=2\ee and
the algebra is $A_{7}^{(2)\wedge}$

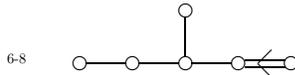
\begin{center}
\scalebox{.5}{
\begin{picture}(180,60)
\put(-45,10){6-8}
\thicklines \multiput(10,10)(40,0){5}{\circle{10}}
\multiput(15,10)(40,0){3}{\line(1,0){30}}
\put(135,7.5){\line(1,0){30}}\put(135,12.5){\line(1,0){30}}
\put(145,10){\line(1,1){10}} \put(145,10){\line(1,-1){10}}
\put(90,50){\circle{10}} \put(90,15){\line(0,1){30}}
\end{picture}
} \end{center}

Case D.3. describes the general structure below in which $\alpha_2$
and $\alpha_3$ have two links while $\alpha_4$ is connected three times

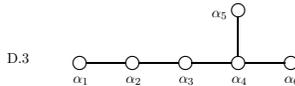
\begin{center}
\scalebox{.5}{
\begin{picture}(180,60)
\put(-45,10){D.3}
\thicklines \multiput(10,10)(40,0){5}{\circle{10}}
\multiput(15,10)(40,0){4}{\line(1,0){30}}
\put(5,-5){$\alpha_1$} \put(45,-5){$\alpha_2$}
 \put(125,-5){$\alpha_4$}  \put(110,45){$\alpha_5$}\put(85,-5){$\alpha_3$}
  \put(165,-5){$\alpha_6$}
\put(130,50){\circle{10}} \put(130,15){\line(0,1){30}}
\end{picture}
} \end{center}

There are two hyperbolic algebras of that type but only one satisfies all
billiard conditions. Its non zero couplings are

\be\lambda = \sqrt{2}\quad\mbox{and}\quad\lambda' =
\mu' = \mu'' = \nu'' =  \nu''' = \rho''' = \rho''''=
1/  \sqrt{2}.\ee
The Dynkin
diagram is

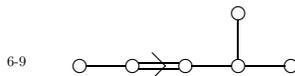
\begin{center}
\scalebox{.5}{
\begin{picture}(180,60)
\put(-45,10){6-9}
\thicklines \multiput(10,10)(40,0){5}{\circle{10}}
\multiput(95,10)(40,0){2}{\line(1,0){30}}
\put(15,10){\line(1,0){30}}
\put(55,7.5){\line(1,0){30}}\put(55,12.5){\line(1,0){30}}
\put(65,0){\line(1,1){10}} \put(65,20){\line(1,-1){10}}
\put(130,50){\circle{10}} \put(130,15){\line(0,1){30}}
\end{picture}
} \end{center}

\textbf{Cases E.}  This set provides all linear diagrams. There
are seven hyperbolic algebras of this kind and all of them are
admissible

\begin{center}
\scalebox{.5}{
\begin{picture}(180,60)
\put(-45,10){E}
\thicklines \multiput(10,10)(40,0){6}{\circle{10}}
\multiput(15,10)(40,0){5}{\line(1,0){30}}
\put(5,-5){$\alpha_1$} \put(45,-5){$\alpha_2$}
 \put(125,-5){$\alpha_4$}  \put(205,-5){$\alpha_6$}\put(85,-5){$\alpha_3$}
  \put(165,-5){$\alpha_5$}
\end{picture}
} \end{center}

E.1. has the following couplings

\begin{eqnarray}
\lambda &=& \sqrt{2} \ ; \
\lambda' = {1 \over \sqrt{ 2}}  \ ; \ \mu' =   \sqrt{ { 3 \over 2}}
\ ; \
\lambda''= 0 \ ; \ \mu'' = \sqrt{ { 2 \over 3}} \ ; \ \nu'' = {2
\over
\sqrt{ 3}} \ ; \
\lambda''' = 0; \nonumber \\ \mu''' &=& 0\ ; \ \nu''' =\sqrt{ 3}  \ ;
\
\rho'''=1 \ ; \
\lambda'''' = 0  \ ; \ \mu''''=0   \ ; \ \nu''''=0\  \ ; \
\rho''''=2
\end{eqnarray} and its Dynkin diagram belongs to $E_6^{(2)\wedge}$

\begin{center}
\scalebox{.5}{
\begin{picture}(180,60)
\put(-45,10){6-10}
\thicklines \multiput(10,10)(40,0){6}{\circle{10}}
\multiput(15,10)(40,0){3}{\line(1,0){30}}
\put(175,10){\line(1,0){30}}
\put(135,7.5){\line(1,0){30}}\put(135,12.5){\line(1,0){30}}
\put(145,10){\line(1,1){10}} \put(145,10){\line(1,-1){10}}
\end{picture}
} \end{center}

E.2. corresponds to

\begin{eqnarray}
\lambda &=&  \sqrt{2} \ ; \
\lambda' = {1 \over \sqrt{ 2}}  \ ; \ \mu' =  \sqrt{{3\over 2}} \ ; \
\lambda''= 0=\ ; \ \mu'' = \sqrt{{2\over 3}}  \ ; \ \nu'' =  { 1 \over
\sqrt{3}} \ ; \
\lambda''' = 0 \nonumber\\ \mu''' &=&0 \ ; \ \nu''' = { \sqrt{3} \over
2}
\ ;
\ \rho'''={1 \over 2} \ ; \
\lambda'''' = 0 \ ; \ \mu''''=0  \ ; \ \nu''''=0 \ ; \
\rho''''=1
\end{eqnarray}
and its algebra is associated to
\begin{center}
\scalebox{.5}{
\begin{picture}(180,60)
\put(-45,10){6-11}
\thicklines \multiput(10,10)(40,0){6}{\circle{10}}
\multiput(15,10)(40,0){2}{\line(1,0){30}}
\multiput(135,10)(40,0){2}{\line(1,0){30}}
\put(95,7.5){\line(1,0){30}}\put(95,12.5){\line(1,0){30}}
\put(105,0){\line(1,1){10}} \put(105,20){\line(1,-1){10}}
\end{picture}
} \end{center}

E.3. The walls are defined through the following set of parameters

\begin{eqnarray}
\lambda &=&  \sqrt{2} \ ; \
\lambda' = {1 \over \sqrt{ 2}}  \ ; \ \mu' =  \sqrt{{3\over 2}} \ ; \
\lambda''= 0\quad \ ; \ \mu'' = \sqrt{{2\over 3}}  \ ; \ \nu'' =  { 2 \over
\sqrt{3}} \ ; \
\lambda'''= 0 \nonumber\\ \mu''' &=&0 \ ; \ \nu''' = { \sqrt{3} \over 2}
\ ;
\ \rho'''={1 \over 2} \ ; \
\lambda'''' = 0 \ ; \ \mu''''=0 \ ; \ \nu''''=0 \ ; \
\rho''''=1
\end{eqnarray} the algebra is $F_4^{\wedge \wedge}$

\begin{center}
\scalebox{.5}{
\begin{picture}(180,60)
\put(-45,10){6-12}
\thicklines \multiput(10,10)(40,0){6}{\circle{10}}
\multiput(15,10)(40,0){3}{\line(1,0){30}}
\put(175,10){\line(1,0){30}}
\put(135,7.5){\line(1,0){30}}\put(135,12.5){\line(1,0){30}}
\put(145,0){\line(1,1){10}} \put(145,20){\line(1,-1){10}}
\end{picture}
} \end{center}

E.4. has the following couplings

\begin{eqnarray}
\lambda &=&  \sqrt{2} \ ; \
\lambda' = {1 \over \sqrt{ 2}} \ ; \ \mu' = {1 \over \sqrt{ 2}} \ ; \
\lambda''= 0 =\ ; \ \mu'' = {1 \over \sqrt{ 2}}   \ ; \ \nu'' =  {1
\over \sqrt{ 2}} \ ; \
\lambda''' = 0 \nonumber \\ \mu''' &=& 0 \ ; \ \nu''' = {1 \over \sqrt{
2}}
\ ;
\ \rho'''={1 \over \sqrt{ 2}}\ ; \
\lambda'''' = 0 \ ; \ \mu''''=0  \ ; \ \nu''''=0 \ ; \
\rho''''=\sqrt{2}  \label{574}
\end{eqnarray}
and its diagram corresponds to $C_4^{\wedge\wedge}$

\begin{center}
\scalebox{.5}{
\begin{picture}(180,60)
\put(-45,10){6-13}
\thicklines \multiput(10,10)(40,0){6}{\circle{10}}
\multiput(95,10)(40,0){2}{\line(1,0){30}}
\put(15,10){\line(1,0){30}}
\put(55,7.5){\line(1,0){30}}\put(55,12.5){\line(1,0){30}}
\put(65,0){\line(1,1){10}} \put(65,20){\line(1,-1){10}}
\put(175,7.5){\line(1,0){30}}\put(175,12.5){\line(1,0){30}}
\put(185,10){\line(1,1){10}} \put(185,10){\line(1,-1){10}}
\end{picture}
} \end{center}

E.5. has the same couplings as those given in (\ref{574}) except
\be \rho'''' = {1\over \sqrt{2}}.\ee Its diagram corresponds to
$A_{8}^{(2)\wedge}$

\begin{center}
\scalebox{.5}{
\begin{picture}(180,60)
\put(-45,10){6-14}
\thicklines \multiput(10,10)(40,0){6}{\circle{10}}
\multiput(95,10)(40,0){2}{\line(1,0){30}}
\put(15,10){\line(1,0){30}}
\put(55,7.5){\line(1,0){30}}\put(55,12.5){\line(1,0){30}}
\put(65,0){\line(1,1){10}} \put(65,20){\line(1,-1){10}}
\put(175,7.5){\line(1,0){30}}\put(175,12.5){\line(1,0){30}}
\put(185,0){\line(1,1){10}} \put(185,20){\line(1,-1){10}}
\end{picture}
} \end{center}

E.6. is characterized by

\begin{eqnarray}
\lambda &=&  \sqrt{2} \ ; \
\lambda' = \sqrt{2} \ ; \ \mu' = \sqrt{2} \ ; \
\lambda''= 0 \ ; \ \mu'' = \sqrt{2} \,  \ ; \ \nu'' =  \sqrt{2}
\ ; \
\lambda''' = 0; \nonumber\\ \mu''' &=& 0 \ ; \ \nu''' = \sqrt{2}   \ ; \
\rho'''=\sqrt{2} \ ; \
\lambda'''' = 0 \ ; \ \mu''''=0  \ ; \ \nu''''=0 \ ; \
\rho''''=\sqrt{2}
\end{eqnarray}
and its diagram describes $D_5^{(2)\wedge}$
\begin{center}
\scalebox{.5}{
\begin{picture}(180,60)
\put(-45,10){6-15}
\thicklines \multiput(10,10)(40,0){6}{\circle{10}}
\multiput(95,10)(40,0){2}{\line(1,0){30}}
\put(15,10){\line(1,0){30}}
\put(55,7.5){\line(1,0){30}}\put(55,12.5){\line(1,0){30}}
\put(65,10){\line(1,1){10}} \put(65,10){\line(1,-1){10}}
\put(175,7.5){\line(1,0){30}}\put(175,12.5){\line(1,0){30}}
\put(185,0){\line(1,1){10}} \put(185,20){\line(1,-1){10}}
\end{picture}
} \end{center}

E.7. is the last one of this rank; its couplings are

\begin{eqnarray}
\lambda &=&  1 \ ; \
\lambda' = {1 \over 2}  \ ; \ \mu' =  {\sqrt{3} \over 2} \ ; \
\lambda''= 0 \ ; \ \mu'' = {1 \over \sqrt{ 3}} \,\, \ ; \ \nu'' =
\sqrt{ {2 \over 3}} \ ; \
\lambda''' = 0; \nonumber \\ \mu''' &=&0 \ ; \ \nu''' = { \sqrt{3} \over 2
\sqrt{2 }}  \ ; \ \rho'''={1 \over 2 \sqrt{ 2}} \ ; \
\lambda'''' = 0 \ ; \ \mu''''=0 \ ; \ \nu''''=0 \ ; \
\rho''''={1 \over \sqrt{ 2}}
\end{eqnarray}
and its diagram gives $A_{8}^{(2)\prime\wedge}$
\begin{center}
\scalebox{.5}{
\begin{picture}(180,60)
\put(-45,10){6-16}
\thicklines \multiput(10,10)(40,0){6}{\circle{10}}
\multiput(55,10)(40,0){2}{\line(1,0){30}}
\put(175,10){\line(1,0){30}} \put(15,12.5){\line(1,0){30}}
\put(15,7.5){\line(1,0){30}} \put(25,0){\line(1,1){10}}
\put(25,20){\line(1,-1){10}}
\put(135,7.5){\line(1,0){30}}\put(135,12.5){\line(1,0){30}}
\put(145,0){\line(1,1){10}} \put(145,20){\line(1,-1){10}}
\end{picture}
} \end{center}

\subsection{$D>3$}

Our next task is again to study which of the 16 algebras admitting
a three-dimensional billiard model allow in addition a higher
dimensional Lagrangian description.

\begin{enumerate}

\item{Diagram $(6-1)$} is the overextension $A_4^{\wedge\wedge}$. The
maximal oxidation dimension is $D_{max}=7$ where the Lagrangian
describes pure gravity \cite{DHJN}.
\item{Diagram $(6-2)$} : Here, $D_{max}=5$. The dominant
walls are the symmetry walls
$\alpha_1=\beta^4-\beta^3$,
$\alpha_2=\beta^3-\beta^2$, $\alpha_3=\beta^2-\beta^1$ and \begin{eqnarray}
\alpha_4 &=&\beta^1 - 1/ \sqrt{3} \phi,\\ \alpha_5 &=&\beta^1 + \beta^2
+ 1/ \sqrt{3} \phi - \psi,\\ \alpha_6 &=&\psi.\end{eqnarray} These are
respectively the electric walls of a one-form, a two-form and a
zero-form. One easily checks that $\tilde\alpha_4=\alpha_5+\alpha_6$,
$\tilde\alpha_5=\alpha_4+\alpha_6$ and $\tilde\alpha_6=
\alpha_2+\alpha_3+\alpha_4+\alpha_5$.
\item{Diagram $(6-3)$} is the overextension $D_4^{\wedge\wedge}$; its 3-D
version can be oxidized up to $D_{max}=6$ and the Lagrangian is
written in references \cite{CJLP} and \cite{DdBHS}.
\item{Diagrams $(6-4)$, $(6-5)$ and $(6-6)$} : their Lagrangians have no
higher dimensional parent.
\item{Diagram $(6-7)$} is the overextension $B_4^{\wedge\wedge}$. Remark
that since the diagram has a fork one can oxidize in two different
ways: both lead to $D_{max}=6$. The Lagrangians can again be found
in references \cite{CJLP} and \cite{DdBHS}.
\item{Diagram $(6-8)$} is the twisted overextension
$A_{7}^{(2)\wedge}$.
$D_{max}=6$. The dominant walls are the symmetry walls
$\alpha_1 = \beta^5-\beta^4$, $\alpha_2 = \beta^4-\beta^3$,
$\alpha_3=\beta^3-\beta^2$ and $\alpha_4 =\beta^2-\beta^1$ and
\begin{eqnarray} \alpha_5 &=&\beta^1+\beta^2 -\phi,\\ \alpha_6
&=&2\phi\end{eqnarray} which are the electric walls of a $2$-form
and a $0$-form. Their respective magnetic walls are
subdominant: indeed one finds \begin{eqnarray}\tilde\alpha_5 &=
&\beta^1+\beta^2+\phi = \alpha_5+\alpha_6\\ \tilde\alpha_6 &=&
\beta^1+\beta^2+\beta^3+\beta^4-2\phi = 2\alpha_5 +
\alpha_4 +2\alpha_3+\alpha_2.\end{eqnarray}
\item{Diagram $(6-9)$} : $D_{max}=4$. The
wall system reads $\alpha_1 = \beta^3-\beta^2,
\alpha_2 =\beta^2-\beta^1$ and
\begin{eqnarray} \alpha_3 &=& \beta^1- 1/ \sqrt{2} \phi,\\
\alpha_4 &=&1/ \sqrt{2} (\phi-\psi), \\ \alpha_5 &=&1/ \sqrt{2}
(\psi-\chi)\\ \alpha_6 &=&1/ \sqrt{2} (\psi+\chi); \end{eqnarray} the last
four are the electric walls of a $1$-form and three $0$-forms. The
subdominant condition is fulfilled: indeed, one finds $\tilde\alpha_3 =
\alpha_3+2\alpha_4+\alpha_5+\alpha_6$, $\tilde\alpha_4 =
\alpha_2+2\alpha_3+\alpha_4+\alpha_5+\alpha_6$, $\tilde\alpha_5 =
\alpha_2+2\alpha_3+2\alpha_4+\alpha_6$, $\tilde\alpha_6 =
\alpha_2+2\alpha_3+2\alpha_4+\alpha_5$.
\item{Diagram $(6-10)$} is the twisted overextension $E_6^{(2)\wedge}$.
The oxidation rule gives the maximal dimension $D_{max}=5$. The walls
other than the symmetry ones are
\begin{eqnarray}\alpha_4 &=&
\beta^1-2/
\sqrt{3}
\phi, \\ \alpha_5 &=& \sqrt{3}\phi - \varphi \\ \alpha_6 &=& 2
\varphi.\end{eqnarray} One checks that $\tilde\alpha_4 =\alpha_3+
2\alpha_4+2\alpha_5+\alpha_6$, $\tilde\alpha_5 = \alpha_2+2\alpha_3+
3\alpha_4+\alpha_5+\alpha_6$, $\tilde\alpha_6 = \alpha_2+2\alpha_3+
3\alpha_4+2\alpha_5$.
\item{Diagram $(6-11)$} : a Lagrangian exists in $D_{max}=5 $ which
produces besides the symmetry walls
\begin{eqnarray}\alpha_4 &=&
\beta^1-1/
\sqrt{3} \phi, \\ \alpha_5 &=& \sqrt{3} /2 \phi - 1/2 \varphi\\
\alpha_6 &=&  \varphi.\end{eqnarray} The subdominant conditions read
$\tilde\alpha_4= \alpha_3+2\alpha_4+2\alpha_5+\alpha_6$, $\tilde\alpha_5=
\alpha_2+2\alpha_3+3\alpha_4+\alpha_5+\alpha_6$ and $\tilde\alpha_6=
\alpha_2+2\alpha_3+3\alpha_4+2\alpha_5$.
\item{Diagram $(6-12)$} is the overextension $F_4^{\wedge\wedge}$;
the maximally oxidized theory is 6 dimensional and contains the
metric, one dilaton, one zero-form, two one-forms, a two-form and
a self- dual three-form field strength \cite{CJLP, DdBHS}.
\item{Diagram $(6-13)$} is the overextension $C_4^{\wedge\wedge}$
\cite{CJLP}. This is the last one of its series: remember that the
$C_n^{\wedge\wedge}$ algebras are hyperbolic only for $n\leq 4$.
The maximal oxidation dimension is $D_{max}=4$; besides the
symmetry walls, the other dominant ones are
\begin{eqnarray}
\alpha_3 &=&
\beta^1-1/
\sqrt{2}
\phi,\\ \alpha_4 &=& 1/ \sqrt{2}( \phi - \varphi), \\ \alpha_5 &=& 1/
\sqrt{2}( \varphi-\psi) \\ \alpha_6 &=& \sqrt{2} \psi.\end{eqnarray} The
subdominant conditions are satisfied, they read
$\tilde\alpha_3 =
\alpha_3+2\alpha_4+2\alpha_5+\alpha_6$, $\tilde\alpha_4 =\alpha_2+
2\alpha_3+\alpha_4+2\alpha_5+\alpha_6$, $\tilde\alpha_5 =
\alpha_2+2\alpha_3+2\alpha_4+\alpha_5+\alpha_6$, $\tilde\alpha_6 =
\alpha_2+2\alpha_3+2\alpha_4+2\alpha_5$.
\item{Diagram $(6-14)$} is the twisted overextension
$A_{8}^{(2)\wedge}$. There is no higher dimensional theory.
\item{Diagram $(6-15)$} represents $D_5^{(2)\wedge}$. In $D_{max}=4$, the
dominant walls other than the symmetry ones are
given by \begin{eqnarray}\alpha_3 &=&
\beta^1-1/
\sqrt{2}
\phi, \\ \alpha_4 &=& 1/ \sqrt{2}( \phi - \varphi), \\ \alpha_5 &=& 1/
\sqrt{2}( \varphi-\psi)\\\alpha_6 &=& 1/ \sqrt{2} \psi.\end{eqnarray} One
obtains easily the following expressions $\tilde\alpha_3 =
\alpha_3+2\alpha_4+2\alpha_5+\alpha_6$, $\tilde\alpha_4 =\alpha_2+
2\alpha_3+\alpha_4+2\alpha_5+2\alpha_6$, $\tilde\alpha_5 =
\alpha_2+2\alpha_3+2\alpha_4+\alpha_5+2\alpha_6$, $\tilde\alpha_6 =
\alpha_2+2\alpha_3+2\alpha_4+2\alpha_5+\alpha_6$.
\item{Diagram $(6-16)$} describes $A_{8}^{(2)\prime\wedge}$; it
cannot be associated to a billiard in
$D>3$.
\end{enumerate}

\textbf{Comment}\newline

Here again, in $D>3$, the subdominant conditions are
always satisfied; it is only in $D=3$ that their r\^ole is crucial in
the selection of the admissible algebras. Hence, they do not add any
constraint in the oxidation construction.

\section{Rank 7, 8, 9 and 10 Hyperbolic algebras}

These hyperbolic algebras fall into two classes: the first one
comprises all algebras of rank $7 \leq r\leq 10$ that are overextensions of the
following finite simple Lie algebras $A_n, \, B_n, \,D_n, E_6, E_7, E_8$. They
are\newline

$A_n^{\wedge\wedge},\quad (n=5,6,7)$

\begin{center}
\scalebox{.5} {
\begin{picture}(180,60)
\put(5,-5){$\alpha_{-1}$} \put(45,-5){$\alpha_0$}
 \put(125,-5){$\alpha_2$}  \put(65,45){$\alpha_n$}\put(85,-5){$\alpha_1$}
  \put(140,45){$\alpha_3$}
\thicklines \multiput(10,10)(40,0){4}{\circle{10}}
\multiput(15,10)(40,0){3}{\line(1,0){30}}
\multiput(90,50)(40,0){2}{\circle{10}}
\put(130,15){\line(0,1){30}} \put(50,15){\line(1,1){35}}
\dashline[0]{2}(95,50)(105,50)(115,50)(125,50)
\end{picture}
}
\end{center}

$B_n^{\wedge\wedge},\quad (n= 5,6,7,8)$

\begin{center}
\scalebox{.5}{
\begin{picture}(180,60)
\put(5,-5){$\alpha_{-1}$} \put(45,-5){$\alpha_0$}
 \put(125,-5){$\alpha_{n-1}$}  \put(70,45){$\alpha_2$}\put(85,-5){$\alpha_1$}
  \put(165,-5){$\alpha_n$}
\thicklines \multiput(10,10)(40,0){5}{\circle{10}}
\multiput(15,10)(40,0){2}{\line(1,0){30}}
\dashline[0]{2}(95,10)(105,10)(115,10)(125,10)
\put(135,7.5){\line(1,0){30}}\put(135,12.5){\line(1,0){30}}
\put(145,0){\line(1,1){10}} \put(145,20){\line(1,-1){10}}
\put(90,50){\circle{10}} \put(90,15){\line(0,1){30}}
\end{picture}
} \end{center}

$D_n^{\wedge\wedge}, \quad (n= 5,6,7,8)$

\begin{center}
\scalebox{.5}{
\begin{picture}(180,60)
\put(5,-5){$\alpha_{-1}$} \put(45,-5){$\alpha_0$}
 \put(125,-5){$\alpha_{n-2}$}  \put(70,45){$\alpha_2$}\put(85,-5){$\alpha_1$}
  \put(165,-5){$\alpha_n$} \put(140,45){$\alpha_{n-1}$}
\thicklines \multiput(10,10)(40,0){4}{\circle{10}}
\multiput(15,10)(40,0){2}{\line(1,0){30}}
\dashline[0]{2}(95,10)(105,10)(115,10)(125,10)
\put(90,50){\circle{10}} \put(90,15){\line(0,1){30}}
\put(130,50){\circle{10}} \put(130,15){\line(0,1){30}}
\multiput(130,10)(40,0){2}{\circle{10}}
\put(135,10){\line(1,0){30}}
\end{picture}
} \end{center}

$E_6^{\wedge \wedge}$

\begin{center}
\scalebox{.5}{
\begin{picture}(180,60)
\put(5,-5){$\alpha_{-1}$} \put(45,-5){$\alpha_0$}
\put(85,-5){$\alpha_1$}
 \put(125,-5){$\alpha_2$}
  \put(165,-5){$\alpha_3$} \put(205,-5){$\alpha_4$}
  \put(140,45){$\alpha_5$}   \put(140,85){$\alpha_6$}
\thicklines \multiput(10,10)(40,0){6}{\circle{10}}
\multiput(15,10)(40,0){5}{\line(1,0){30}}
\put(130,50){\circle{10}} \put(130,15){\line(0,1){30}}
\put(130,90){\circle{10}} \put(130,55){\line(0,1){30}}
\end{picture}
} \end{center}

$E_7^{\wedge \wedge}$

\begin{center}
\scalebox{.5}{
\begin{picture}(180,60)
\put(5,-5){$\alpha_{-1}$} \put(45,-5){$\alpha_0$}
\put(85,-5){$\alpha_1$}
 \put(125,-5){$\alpha_2$}
  \put(165,-5){$\alpha_3$} \put(205,-5){$\alpha_4$}
  \put(245,-5){$\alpha_5$}   \put(285,-5){$\alpha_6$}
  \put(180,45){$\alpha_7$}
\thicklines \multiput(10,10)(40,0){8}{\circle{10}}
\multiput(15,10)(40,0){7}{\line(1,0){30}}
\put(170,50){\circle{10}} \put(170,15){\line(0,1){30}}
\end{picture}
} \end{center}

$E_8^{\wedge \wedge}$

\begin{center}
\scalebox{.5}{
\begin{picture}(180,60)
\put(5,-5){$\alpha_{-1}$} \put(45,-5){$\alpha_0$}
\put(85,-5){$\alpha_1$}
 \put(125,-5){$\alpha_2$}
  \put(165,-5){$\alpha_3$} \put(205,-5){$\alpha_4$}
  \put(245,-5){$\alpha_5$}   \put(285,-5){$\alpha_6$}
  \put(325,-5){$\alpha_7$}
  \put(260,45){$\alpha_8$}
\thicklines \multiput(10,10)(40,0){9}{\circle{10}}
\multiput(15,10)(40,0){8}{\line(1,0){30}}
\put(250,50){\circle{10}} \put(250,15){\line(0,1){30}}
\end{picture}
} \end{center}

In the second class, one finds the
four duals of
the
$B_n^{\wedge\wedge}, (n= 5,6,7,8)$, i.e. the algebras known as $CE_{n+2} =
A_{2n-1}^{(2)\wedge}$
\newline

$CE_{n+2} = A_{2n-1}^{(2)\wedge}$

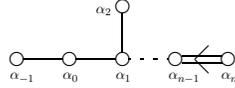
\begin{center}
\scalebox{.5}{
\begin{picture}(180,60)
\put(5,-5){$\alpha_{-1}$} \put(45,-5){$\alpha_0$}
 \put(125,-5){$\alpha_{n-1}$}  \put(70,45){$\alpha_2$}\put(85,-5){$\alpha_1$}
  \put(165,-5){$\alpha_n$}
\thicklines \multiput(10,10)(40,0){5}{\circle{10}}
\multiput(15,10)(40,0){2}{\line(1,0){30}}
\dashline[0]{2}(95,10)(105,10)(115,10)(125,10)
\put(135,7.5){\line(1,0){30}}\put(135,12.5){\line(1,0){30}}
\put(145,10){\line(1,1){10}} \put(145,10){\line(1,-1){10}}
\put(90,50){\circle{10}} \put(90,15){\line(0,1){30}}
\end{picture}
} \end{center}

\subsection{Overextensions of finite simple Lie algebras}
The algebras of the first class have already been encountered as
billiards of some three-dimensional $G/H$ coset theories as
explained in \cite{DdBHS} to which we refer for more information.
Those of rank $10$, $E_{10}, BE_{10}$ and $DE_{10}$ have been
found \cite{DH2} to describe the billiards of the seven string
theories, $M, IIA, IIB, I, HO, HE$ and the closed bosonic string
in 10 dimensions. More precisely, these theories split into three
separate blocks which correspond to three distinct billiards:
namely, ${\cal B}_2 = \{M, IIA, IIB\}$ leads to $E_{10}$, ${\cal
B}_1 = \{I, HO, HE\}$ corresponds to $BE_{10}$ and ${\cal B}_0 =
\{D=10\, \mbox{closed bosonic string}\}$ gives $DE_{10}$.
\newline

For sake of completeness, we here simply recall the maximal
spacetime dimensions and the specific
$p$-forms menus producing
the billiards.

\begin{enumerate}
\item{$A_n^{\wedge \wedge},\quad (n=5,6,7$)} : the Lagrangian is that of pure
gravity in
$D_{max}=n+3$.

\item{$B_n^{\wedge\wedge},\quad (n=5,6,7,8$)} : the maximally oxidized
Lagrangian lives in
$D_{max}= n+2$ where it comprises a dilaton, a $1$-form coupled to the
dilaton with coupling equal to $\lambda^{(1)}(\phi)=\phi/\sqrt{d-1}$ and a
$2$-form coupled to the dilaton with coupling equal to
$\lambda^{(2)}(\phi)=2\phi/ \sqrt{d-1}$.

\item{$D_n^{\wedge \wedge}, \quad (n=5,6,7,8$)} : a Lagrangian exists in
$D_{max}= n+2$ and comprises a dilaton and a $2$-form coupled to
the dilaton with coupling equal to $\lambda^{(2)}(\phi)= 2\phi/ \sqrt{d-1}$.

\item{$E_6^{\wedge \wedge}$} : the maximal oxidation dimension is $D_{max}=8$.
The Lagrangian has a dilaton, a $0$-form with coupling
$\lambda^{(0)}(\phi)=\phi\,\sqrt{2}$ and a
$3$-form with coupling
$\lambda^{(3)}(\phi)=-\phi/
\sqrt{2}$.

\item{$E_7^{\wedge \wedge}$} :  the maximal spacetime
dimension is $D_{max}=10$. The Lagrangian describes gravity and a
$4$-form: it is a truncation of type IIB supergravity.

\item{$E_8^{\wedge \wedge}$} : $D_{max}=11$. The
Lagrangian describes gravity coupled to a $3$-form; it is the bosonic sector
of eleven dimensional supergravity.

\end{enumerate}

\subsection{The algebras $CE_{n+2} = A_{2n-1}^{(2)\wedge}$}

The Weyl chamber of the algebras $CE_{n+2}$ ($n=5,6,7,8$), which
are dual to $B_n^{\wedge\wedge}$, allows a billiard realization in
maximal dimension $D_{max} = n+1 = d+1$. The field content of the
theory is the following: there are two dilatons, $\phi$ and
$\varphi$, a $0$-form coupled to the dilatons through \be
\lambda^{(0)}(\phi) = 2 \sqrt{\frac{(d-1)}{d}}\,\phi - \frac{2}{
\sqrt{d}}\,\varphi,\ee a one form with dilaton couplings \be
\lambda^{(1)}(\phi) = -\sqrt{\frac{d}{(d-1)}}\,\phi\ee and a
2-form with the following couplings \be \lambda^{(2)}(\phi) = -
\frac{2}{ \sqrt{d (d-1)}} \,\phi - \frac{2}{
\sqrt{d}}\,\varphi.\ee

In particular, the Lagrangian in $D_{max}=9$ producing the
billiard identifiable as the fundamental Weyl chamber of $CE_{10}$
corresponds to $n=8=d$ and is explicitly given by \cite{MHBJ}

\begin{eqnarray} {\cal L}_{9} &=& ^{(9)}R\star\unity - \star
d\phi\wedge d\phi - \star d\varphi \wedge d\varphi
-\frac{1}{2}\,e^{(2\phi\sqrt{\frac{7}{2}} - \varphi\sqrt{2} \,)}\star
F^{(1)}\wedge F^{(1)}\nonumber
\\ & \, & -\frac{1}{2}\,e^{-4\phi\sqrt{\frac{2}{7}}}\star
F^{(2)}\wedge
F^{(2)}-\frac{1}{2}\,e^{-(\phi\sqrt{\frac{2}{7}}+\varphi\sqrt{2}\,)}\star
F^{(3)}\wedge F^{(3)}.
\label{Lagr}\end{eqnarray}

$CE_{10}$ is the fourth hyperbolic algebra of rank 10; contrary to
the other three cited above, which belong to the class of the
overtextensions of finite simple Lie algebras, its Lagrangian
(\ref{Lagr}) does not stem from string theories.\newline

\section{Conclusions}

In this paper we have presented all Lagrangian systems in which
gravity, dilatons and $p$-forms combine in such a way as to
produce a billiard that can be identified with the Weyl chamber of
a given hyperbolic Kac Moody algebra. Exhaustive results have been
systematically obtained by first constructing Lagrangians in three
spacetime dimensions, at least for the algebras of rank $r\leq 6$.
We insist on the fact that our three-dimensional Lagrangians are
not assumed to realize a coset theory. We also have solved the
oxidation problem and provided the Lagrangians in the maximal
spacetime dimension with their $p$-forms content and specific
dilaton couplings. It turns out that the subdominant conditions
play no r\^ole in the oxidation analysis. The positive integer
coefficients that appear when expressing the subdominant walls in
terms of the dominant ones in the maximal oxidation dimension have
been systematically worked out.
\newline

\section{More hyperbolic algebras}

For completeness, we draw hereafter the Dynkin diagrams of 6
hyperbolic algebras missing in reference \cite{S}. This raises their
total number to $142$.

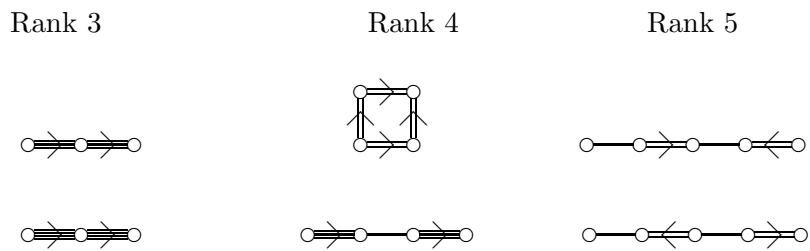
\begin{center}
\begin{tabular}{l c c}
Rank 3 & Rank 4 & Rank 5 \\
& & \\
\scalebox{.5}{
\begin{picture}(180,60)
\thicklines \multiput(10,10)(40,0){3}{\circle{10}}
\put(15,7.5){\line(1,0){30}} \put(15,12.5){\line(1,0){30}}
\put(15,10){\line(1,0){30}}
\put(25,0){\line(1,1){10}} \put(25,20){\line(1,-1){10}}
\put(55,7.5){\line(1,0){30}} \put(55,12.5){\line(1,0){30}}
\put(55,10){\line(1,0){30}}
\put(65,0){\line(1,1){10}} \put(65,20){\line(1,-1){10}}
\end{picture}
}
&
\scalebox{.5}{
\begin{picture}(180,60)
\thicklines \multiput(50,10)(40,0){2}{\circle{10}}
\put(55,8){\line(1,0){30}} \put(55,12){\line(1,0){30}}
\put(50,50){\circle{10}}
\put(55,48){\line(1,0){30}} \put(55,52){\line(1,0){30}}
\put(90,50){\circle{10}}
\put(48,15){\line(0,1){30}} \put(52,15){\line(0,1){30}}
\put(88,15){\line(0,1){30}} \put(92,15){\line(0,1){30}}
\put(65,0){\line(1,1){10}} \put(65,20){\line(1,-1){10}}
\put(65,40){\line(1,1){10}} \put(65,60){\line(1,-1){10}}
\put(40,25){\line(1,1){10}} \put(50,35){\line(1,-1){10}}
\put(80,25){\line(1,1){10}} \put(90,35){\line(1,-1){10}}
\end{picture}
}
&
\scalebox{.5}{
\begin{picture}(180,60)
\thicklines \multiput(10,10)(40,0){5}{\circle{10}}
\put(15,10){\line(1,0){30}} \put(95,10){\line(1,0){30}}
\put(55,8){\line(1,0){30}} \put(55,12){\line(1,0){30}}
\put(65,0){\line(1,1){10}} \put(65,20){\line(1,-1){10}}
\put(135,8){\line(1,0){30}} \put(135,12){\line(1,0){30}}
\put(145,10){\line(1,1){10}} \put(145,10){\line(1,-1){10}}
\end{picture}
}
  \\
\scalebox{.5}{
\begin{picture}(180,60)
\thicklines \multiput(10,10)(40,0){3}{\circle{10}}
\put(15,8.75){\line(1,0){30}} \put(15,11.25){\line(1,0){30}}
\put(15,6.25){\line(1,0){30}} \put(15,13.75){\line(1,0){30}}
\put(25,0){\line(1,1){10}} \put(25,20){\line(1,-1){10}}
\put(55,8.75){\line(1,0){30}} \put(55,11.25){\line(1,0){30}}
\put(55,6.25){\line(1,0){30}} \put(55,13.75){\line(1,0){30}}
\put(65,0){\line(1,1){10}} \put(65,20){\line(1,-1){10}}
\end{picture}
} &

\scalebox{.5}{
\begin{picture}(180,60)
\thicklines \multiput(10,10)(40,0){4}{\circle{10}}
\put(15,10){\line(1,0){30}} \put(15.5,12.5){\line(1,0){30}}
\put(15,7.5){\line(1,0){30}} \put(25,0){\line(1,1){10}}
\put(25,20){\line(1,-1){10}}
\put(55,10){\line(1,0){30}}
\put(95,12.5){\line(1,0){30}} \put(95,7.5){\line(1,0){30}}
\put(95,10){\line(1,0){30}} \put(105,0){\line(1,1){10}}
\put(105,20){\line(1,-1){10}}
\end{picture}
}
&
\scalebox{.5}{
\begin{picture}(180,60)
\thicklines \multiput(10,10)(40,0){5}{\circle{10}}
\put(15,10){\line(1,0){30}} \put(95,10){\line(1,0){30}}
\put(55,8){\line(1,0){30}} \put(55,12){\line(1,0){30}}
\put(65,10){\line(1,1){10}} \put(65,10){\line(1,-1){10}}
\put(135,8){\line(1,0){30}} \put(135,12){\line(1,0){30}}
\put(145,0){\line(1,1){10}} \put(145,20){\line(1,-1){10}}
\end{picture}}
\\
\end{tabular}
\end{center}

\section{Acknowledgements}

We are grateful to Marc Henneaux for suggesting the problem. This
work is supported in part by the ``Actions de Recherche
Concert{\'e}es" of the ``Direction de la Recherche Scientifique -
Communaut{\'e} Fran{\c c}aise de Belgique", by a ``P\^ole
d'Attraction Interuniversitaire" (Belgium), by IISN-Belgium
(convention 4.4505.86). Support from the European Commission RTN
programme HPRN-CT-00131, in which we are associated to K. U.
Leuven, is also acknowledged.

\end{document}